\documentclass[12pt]{article}
\usepackage[dvips]{color}
\usepackage{epsfig}
\usepackage{amsmath}
\usepackage{graphicx}
\usepackage{epstopdf}

\begin{document}
\title{\bf A Universe with a generalized ghost dark energy and Van der Waals fluid interacting with a fluid}
\author{{M. Khurshudyan$^{a,}$\thanks{Email:
martiros.khurshudyan@mpikg.mpg.de}, B. Pourhassan$^{b,}$\thanks{Email: bpourhassan@yahoo.com}}\\
$^{a}${\small {\em Max Planck Institute of Colloids and Interfaces,}}\\
{\small{Potsdam-Golm Science Park Am Muhlenberg 1 OT Golm, 14476 Potsdam}}\\
$^{b}${\small {\em School of Physics, Damghan University, Damghan, Iran}}}  \maketitle
\begin{abstract}
In this paper, we consider an unusual connection between different fluids. Having established a research goal we would like to consider a toy model of the Universe and investigate its behavior, especially for later time evolution for well known facts. The main goal of the article is to consider a toy model of the Universe with generalized ghost dark energy, Van der Waals gas and a phenomenologically modified fluid. The origin of the last component can be understood as a result of interaction between some original fluid and some source of energy or matter in Universe. By unusual connection we mean an assumption that generalized ghost dark energy has its contribution to the model by an interaction term $Q$ and we suppose an interaction of the form $Q=3Hb(\rho_{\small{tot}}-\rho_{GDe})$. Graphical analysis is performed and the questions of validity of the generalized second law of thermodynamics and stability of the model also approached in this paper.
\end{abstract}
\newpage
\section{\large{Introduction}}
The observations of high redshift type SNIa supernovae \cite{Riess}-\cite{Amanullah} reveal the speeding up expansion of our universe, basis and the nature of which is not clear. The surveys of clusters of galaxies show that the density of matter is very much less than critical density \cite{Pope}, observations of Cosmic Microwave Background (CMB) anisotropies indicate that the universe is flat and the total energy density is very close to the critical $\Omega_{\small{tot}} \simeq1$ \cite{Spergel}. Based on experimental data some mysterious component of the energy is thought to be responsible of the physics of the accelerated expansion but it seems that it is not alone in Universe and it has a partner or opponent which is called dark matter. The mysterious component of the energy called dark energy and it is described by negative pressure and positive energy density giving negative EoS parameter. According to different estimations Dark energy occupies about 73$\% $ of the energy of our universe, other component, dark matter, about 23$\%$, and usual baryonic matter occupy about 4$\%$. Several different attempts are used to explain now days Universe. Cosmological constant is a simple model of the dark energy. However, in presence of many research articles and research direction questions like the origin of dark energy and dark matter is still unknown, the possible connection between them is unknown, real role of the components to the history of the Universe and physical processes are unknown still. This situation, unfortunately or fortunately, gives a lot of freedom to researchers and possibility of some simulations. Alternative models of dark energy suggest a dynamical form of dark energy, which at least in an effective level, can originate from a variable cosmological constant \cite{Sola}, or from various fields, such is a canonical scalar field \cite{Ratra}-\cite{Saridakis0} (quintessence), a phantom field, that is a scalar field with a negative sign of the kinetic term \cite{Caldwell}-\cite{Dutta}, or the combination of quintessence and phantom in a unified model named quintom \cite{Feng}-\cite{Qiu}. Finally, an interesting attempt to probe the nature of dark energy according to some basic quantum gravitational principles are the holographic dark energy paradigm \cite{Hsu}-\cite{hde} and agegraphic dark energy models \cite{Cai1}-\cite{WeiCai2}. Alternative models for cosmological constant were considered, because with cosmological constant we faced with two problems i.e. absence of a fundamental mechanism which sets the cosmological constant zero or very small value the problem known as fine-tuning problem, because in the framework of quantum field theory, the expectation value of vacuum energy is 123 order of magnitude larger than the observed value \cite{Steinhardt}. The second problem known as cosmological coincidence problem, which asks why are we living in an epoch in which the densities of dark energy and matter are comparable? Other interesting way to solve above mentioned problems is to consider interactions between components. Since no known symmetry in nature prevents or suppresses a nonminimal coupling between dark energy and dark matter, there may exist interactions between the two components. At the same time, from observation side, no piece of evidence has been so far presented against such interactions. Indeed, possible interactions between the two dark components have been discussed in recent years. It is found that a suitable interaction can help to alleviate the coincidence problem. Different interacting models of dark energy have been investigated. The dark energy was entered to the general relativity by hand, however some modifications of gravity, like $F(R)$, $F(T)$ or $F(G)$ \cite{Kazuharu} (and references therein) can give a base for the origin of the dark energy and we do not need operate by hand anymore. However, these theories should pass experimental tests, which seems are can not be done in this stage. Theoretically, some forms (functions) of the proposed modifications brings some finite time future singularities, which should be understood not phenomenologically, but in more fundamental level. Considering modified gravities we can understood formally that we have operated with a geometrical part of the action, but action also contains matter part which also could be modified. With this approach in literature we account a tendency to present fluids which have not ordinary EoS equations, they are inhomogeneous and they contain some "strange" unification of different fluids. This approach could be interpreted as a willing to find a one fluid description, where some fluid concerning to the evolution was evolved to some other fluid which we observed today. As was mentioned above there are several thoughts concerning to the dark energy and dark matter, even there is a belief that they are the same operating on different scales. Whatever, now we assume that there is a connection between them, usually called interaction, that allows to transform some dark energy to dark matter and opposite and why today energy densities of both are of the same order. Interaction as a part of the dynamics will play important role and over the years different types of the interaction term was proposed, bellow we will come back to this question also.\\
Our purpose is to consider a scenario where one of the components of the fluids has (had or will have) unusual role in the physics of the Universe. In this stage we thought that it can be through an interaction $Q$. Bellow we presented the basics components of the Universe and assign each of them a role, then we will be back to the question of the interaction and discuss about it also. After all we will start to analyze the model. We consider a Universe composed of Van der Waals fluid \cite{Callen} and \cite{Kremer},
\begin{equation}\label{eq:Waals}
P_{w}=\frac{8\omega_{w}\rho_{w}}{3-\rho_{w}}-3\rho_{w}^{2},
\end{equation}
where $\omega_{w}$ can be associated with a EoS parameter of a barotropic fluid. Van der Waals gas could be accounted as a fluid with unusually EoS as was discussed before or could be thought a fluid satisfying to more general form of EoS i.e. $F(\rho,P)=0$. We suppose it interacting with a modified fluid model \cite{Zimdahl},
\begin{equation}\label{eq:modfluid}
\rho_{m}=\rho_{m0}a^{-3}f(a),
\end{equation}
and,
\begin{equation}\label{eq:f}
f(a)=1+\gamma a^{5}\exp[-a^{2}/\sigma^{2}].
\end{equation}
where $a(t)$ is a scale factor, $\gamma$ is an interaction coefficient. The origin of such fluid can be supposed to be an interaction or different types of interactions between some original fluid and other fluids or resources in Universe, which are coded in $\gamma$. Furthermore, as a different model we could consider more general fluid with the following form
\begin{equation}\label{eq:fmod}
f(a,n)=1+\gamma a^{n}\exp[-a^{2}/\sigma^{2}],
\end{equation}
where we suppose a possibility to have parameter $n$ instead of  number for a corresponding term. In an Appendix, we analyze cosmological parameters of the Universe with above defined form of $f(a,n)$.\\
Among various models of dark energy, a new model of dark energy called Veneziano ghost dark energy of our interest has been recently proposed, which supposed to exist to solve the $U(1)_{A}$ problem in low-energy effective theory of QCD, has attracted a lot of interests in
recent years \cite{Ghost1}-\cite{Chao-Jun3}. Indeed, the contribution of the ghosts field to the vacuum energy in curved space or time-dependent background can be regarded
as a possible candidate for the dark energy. It is completely decoupled from the physics sector. Veneziano ghost is unphysical in the QFT formulation in Minkowski space-time, but exhibits important non trivial physical effects in the expanding Universe and these effects give rise to a vacuum energy density $\rho_{D}\sim \Lambda^{3}_{QCD}H\sim (10^{-3}eV)^{4}$. With $H\sim 10^{-33}eV$ and $\Lambda_{QCD}\sim 100 eV$ we have the right value for the force accelerating the Universe today. It is hard to accept such linear behavior and it is thought that there should be some exponentially small corrections. However, it can be argued that the form of this behavior can be result of the fact of the very complicated topological structure of strongly coupled QCD. This model has advantage compared to other models of dark energy, that it can be explained by standard model and general relativity. Comparison with experimental data reveal that the current data does not favor it compared to the $\Lambda$CDM model, which is not conclusive and future study of the problem is needed. Energy density of ghost dark energy reads as,
\begin{equation}\label{eq:GDE}
\rho_{\small{GDe}}=\alpha H,
\end{equation}
where $H$ is Hubble parameter $H=\dot{a}/a$ and $\alpha$ is constant parameter of the model, which should be determined. A generalization of the model \cite{Cai} also was proposed for which energy density and reads as,
\begin{equation}\label{eq:GDEgen}
\rho_{\small{GDe}}=\alpha H+\beta H^{2},
\end{equation}
with $\alpha$ and $\beta$ constant parameters of the model. Such kind of fluids could be named as a geometrical fluids, because from the description it is clear that it contains information about geometry of the space-time and metric.
Concerning to this component we assume that it will appear and play a role only in the interaction term. Generally what is done in this field is that we assume existence of all fluids in the field equations, but here we suppose that one of the fluids can be an intermediate fluid arose in case of the interaction i.e before transforming to Van der Waals gas our fluid becomes to ghost dark energy and after some time it transforms to Van der Waals gas fully. One of the ways to solve the cosmological coincidence problem discussed above is to consider the interaction between the components (on phenomenological level) as already were mentioned. Interaction could be considered as a function of energy densities and their derivatives: $Q(\rho_{i},\dot{\rho}_{i},\ldots)$. To make everything working with units, we only should consider the fact that unit of interaction term $[Q]=\frac{[energy ~ density]}{time}$ and assume that unit $time^{-1}$ could be contributed, for instance, from Hubble parameter. Over years different models of interactions were proposed and considered, but everything was done on phenomenological level only. Very intensively were considered interactions $Q=3Hb\rho_{m}$, $Q=3Hb\rho_{\small{de}}$, $Q=3Hb\rho_{\small{tot}}$, where $b>0$ is a coupling constant, other form of interactions, where question of time unit were solved with help of first order time derivative $Q=\gamma\dot{\rho}_{\small{m}}$, $Q=\gamma\dot{\rho}_{\small{de}}$, $Q=\gamma\dot{\rho}_{\small{tot}}$. Interaction of the general form  $Q=3Hb\gamma\rho_{i}+\gamma\dot{\rho_{i}}$, where $i=\{m,de,tot\}$ also captured a lot of attention. Interaction between components arose as a result of splitting of the energy conservation, which mathematically can be described as follows,
\begin{equation}
\dot{\rho}_{1}+3H(\rho_{1}+P_{1})=Q,
\end{equation}
and,
\begin{equation}
\dot{\rho}_{2}+3H(\rho_{2}+P_{2})=-Q,
\end{equation}
which could be understood as: there is not energy conservation for the components separately, but due to interaction between all components, energy of whole mixture conserves. This approach is correct for this moment only. Other type of interaction considered in literature and supported by experimental data is sign-changeable interaction \cite{Hao}, \cite{Hao2}, \cite{Sun}. Motivated by \cite{Sun} work, in this stage we would like to consider an interaction where energy density of the Ghost dark energy appears in this form,
\begin{equation}\label{eq:ourint}
Q=3Hb(\rho-\rho_{GDe}).
\end{equation}
The mixture of our consideration will be described by $\rho$ and $P$ given by,
\begin{equation}\label{eq:mixture energy}
\rho = \rho_{ \small{w}}+\rho_{ \small{m}}
\end{equation}
and,
\begin{equation}\label{eq:mixture presure}
P = P_{w}+P_{m}.
\end{equation}
This paper organized as follow, after introduction, in next section we will give field equations and will discuss question of stability of our model. Graphical analysis of different cosmological parameters are discussed as well for the case of ghost dark energy. Then, we study the case of generalized ghost dark energy. In Appendix we discuss question of generalized second law of thermodynamics \cite{1,2,3,4} and cosmological parameters of a general fluid with $f(a,n)$.
\section{Field equations}
\begin{figure}[h!]
 \begin{center}$
 \begin{array}{cccc}
\includegraphics[width=50 mm]{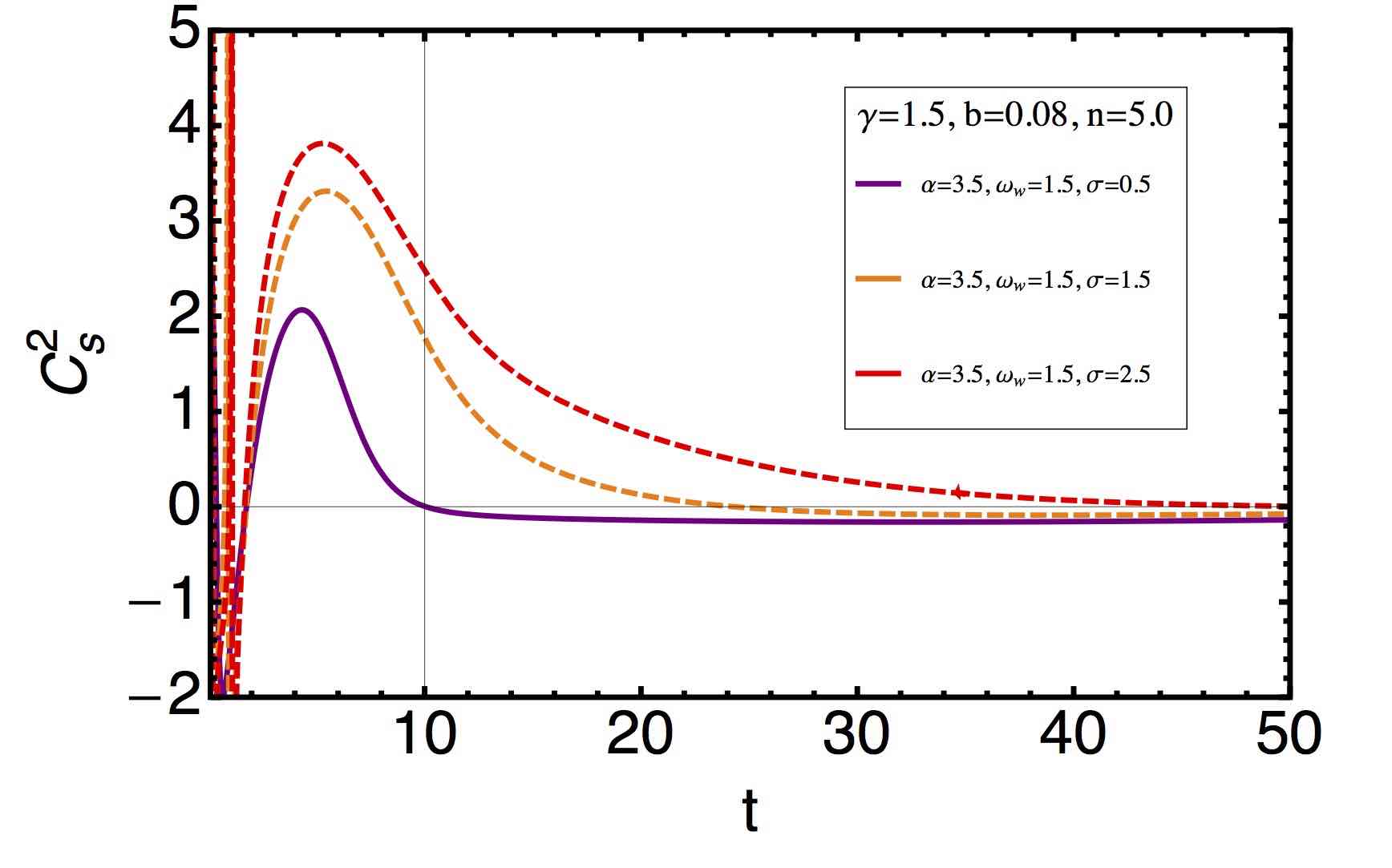} &
\includegraphics[width=50 mm]{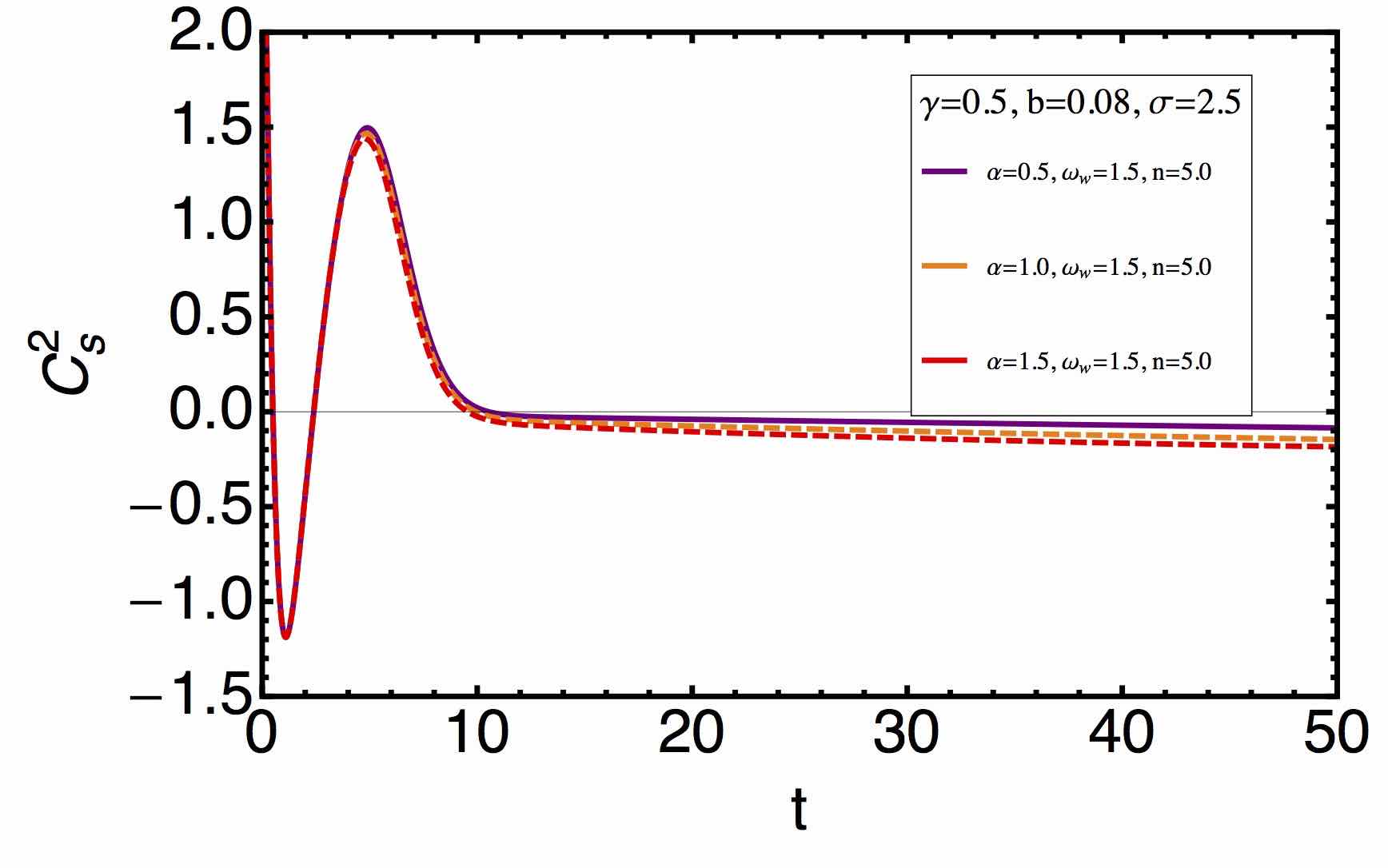}\\
 \end{array}$
 \end{center}
\caption{Behavior of squared sound speed $C_{s}^{2}$ against $t$.}
 \label{fig:0}
\end{figure}

Field equations that govern our model of consideration are,
\begin{equation}\label{eq:Einstein eq}
R^{\mu\nu}-\frac{1}{2}g^{\mu\nu}R^{\alpha}_{\alpha}=T^{\mu\nu}.
\end{equation}
By using the
following FRW metric for a flat Universe,
\begin{equation}\label{s2}
ds^2=-dt^2+a(t)^2\left(dr^{2}+r^{2}d\Omega^{2}\right),
\end{equation}
field equations can be reduced to the following Friedmann equations,
\begin{equation}\label{eq: Fridmman vlambda}
H^{2}=\frac{\dot{a}^{2}}{a^{2}}=\frac{\rho}{3},
\end{equation}
and,
\begin{equation}\label{eq:Freidmann2}
\dot{H}=-\frac{1}{2}(\rho+P),
\end{equation}
where $d\Omega^{2}=d\theta^{2}+\sin^{2}\theta d\phi^{2}$, and $a(t)$
represents the scale factor. The $\theta$ and $\phi$ parameters are
the usual azimuthal and polar angles of spherical coordinates, with
$0\leq\theta\leq\pi$ and $0\leq\phi<2\pi$. The coordinates ($t, r,
\theta, \phi$) are called co-moving coordinates.\\
Energy conservation $T^{;j}_{ij}=0$ reads as,
\begin{equation}\label{eq:Bianchi eq}
\dot{\rho}+3H(\rho+P)=0.
\end{equation}
To introduce an interaction between the dark energy and the dark matter (\ref{eq:Bianchi eq}) we should mathematically split it into two following equations,
\begin{equation}\label{eq:inteqm}
\dot{\rho}_{m}+3H(\rho_{m}+P_{m})=Q,
\end{equation}
and,
\begin{equation}\label{eq:inteqG}
\dot{\rho}_{w}+3H(\rho_{w}+P_{w})=-Q.
\end{equation}
With our assumptions last two equations could be written in a such way that we can obtain pressure of matter and energy density of Van der Waals gas. After some mathematics we will have,
\begin{equation}\label{eq:PM}
P_{m}=Hb(3H-\alpha)-\frac{\rho_{m0}}{3}\gamma a^{n-3}(n-\frac{2a^{2}}{\sigma^{2}})\exp[-a^{2}/\sigma^{2}],
\end{equation}
and,
\begin{equation}\label{eq:inteqG1}
\dot{\rho}_{w}+3H\rho_{w}(1+\omega)+3H^{2}b(3H-\alpha)=0,
\end{equation}
where,
\begin{equation}
\omega=\frac{P_{w}}{\rho_{w}}=\frac{8\omega_{w}}{3-\rho_{w}}-3\rho_{w}.
\end{equation}
The last equation gives us possibility to obtain behavior of $\rho_{w}$. Cosmological parameters of our interest are scale factor $a$, $\rho$, $P$, $\omega_{tot}=\frac{P_{m}+P_{w}}{\rho_{m}+\rho_{w}}$ and deceleration parameter $q$,
\begin{equation}\label{eq:decparameter}
q=-\frac{1}{H^{2}} \frac{\ddot{a}}{a}=-1-\frac{\dot{H}}{H^{2}}.
\end{equation}
Hence, we can obtain energy density and pressure, one can investigate stability
of theory via sound speed,
\begin{equation}
C^{2}_{s}=\frac{\dot{P}}{\dot{\rho}},
\end{equation}
so $C^{2}_{s} \ge 0$ yield to stability of theory. Graphical analysis of $C^{2}_{s}$ shows that our theory could be stable in early epochs, while for later stages, when $\omega_{tot}\rightarrow -1$ it could be unstable $C^{2}_{s} < 0$ depends on values of parameters of the model. However, we obtain that for an intermediate regime, which corresponds to an accelerated expansion of the Universe $q<0$ and $\omega_{tot}>-1$ our theory is stable again with $C^{2}_{s} \ge 0$. First plot in Fig. \ref{fig:0} represent behavior of speed of sound against time as a function of $\sigma$. For a fixed values of other parameters we observe that with increasing value of $\sigma$ we can obtain stable theory for later stages of evolution. For instance with $b=0.08$, $n=5.0$, $\gamma=1.5$, $\alpha=2.5$, $\omega_{w}=1.5$ and $\sigma=2.5$ in fact we have stable theory for a sufficiently long time for evolution even with $\omega_{tot} \rightarrow -1$. Second plot present behavior of  $C^{2}_{s}$ over time as a function of $\alpha$. Analysis shows that with increasing value of $\alpha$ we are increasing unstability of the theory for later stages of evolution. During analysis, for later stages of evolution the same behavior for speed of sound were observed for other parameters of the model.

\begin{figure}[h!]
 \begin{center}$
 \begin{array}{cccc}
\includegraphics[width=50 mm]{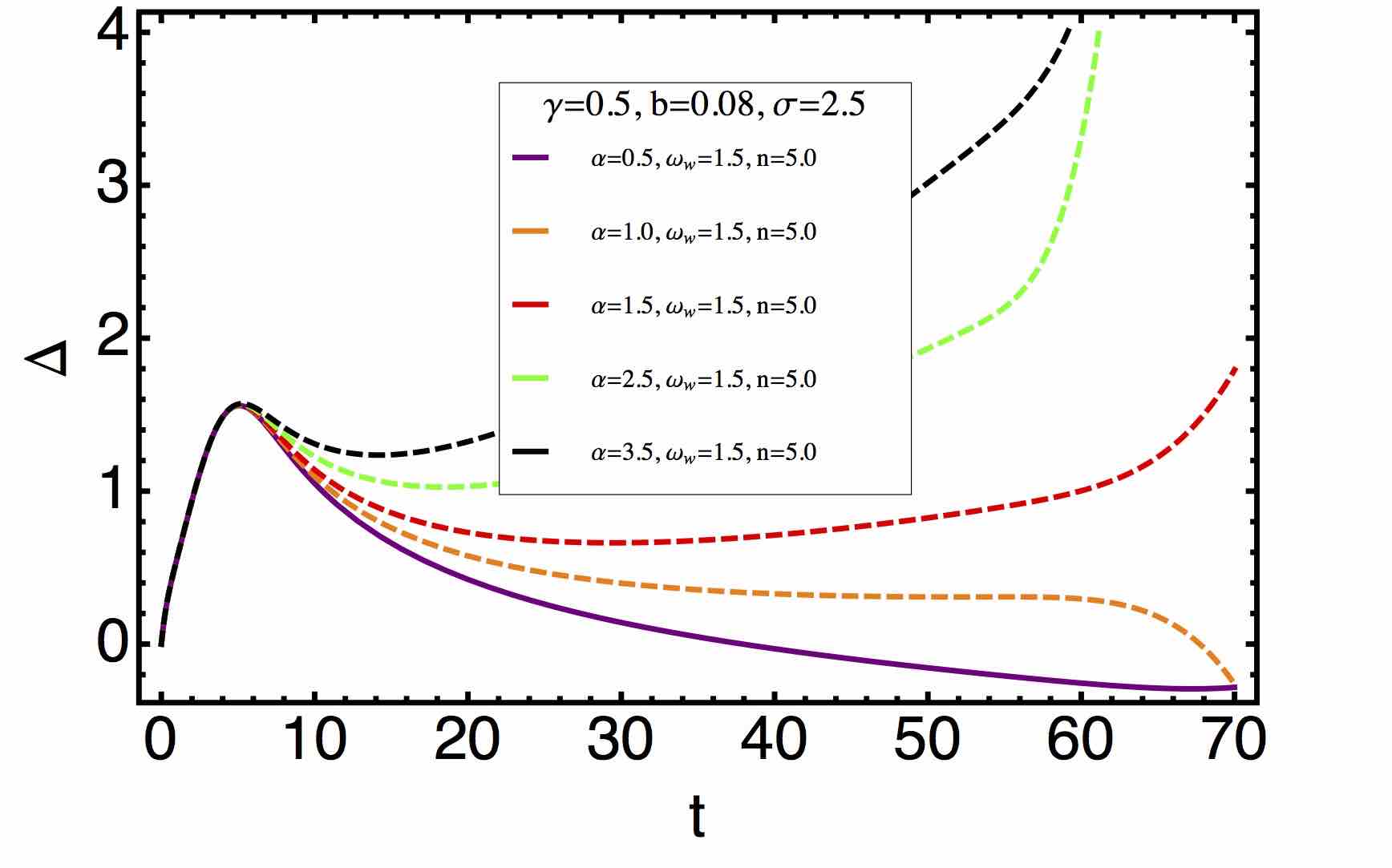} &
\includegraphics[width=50 mm]{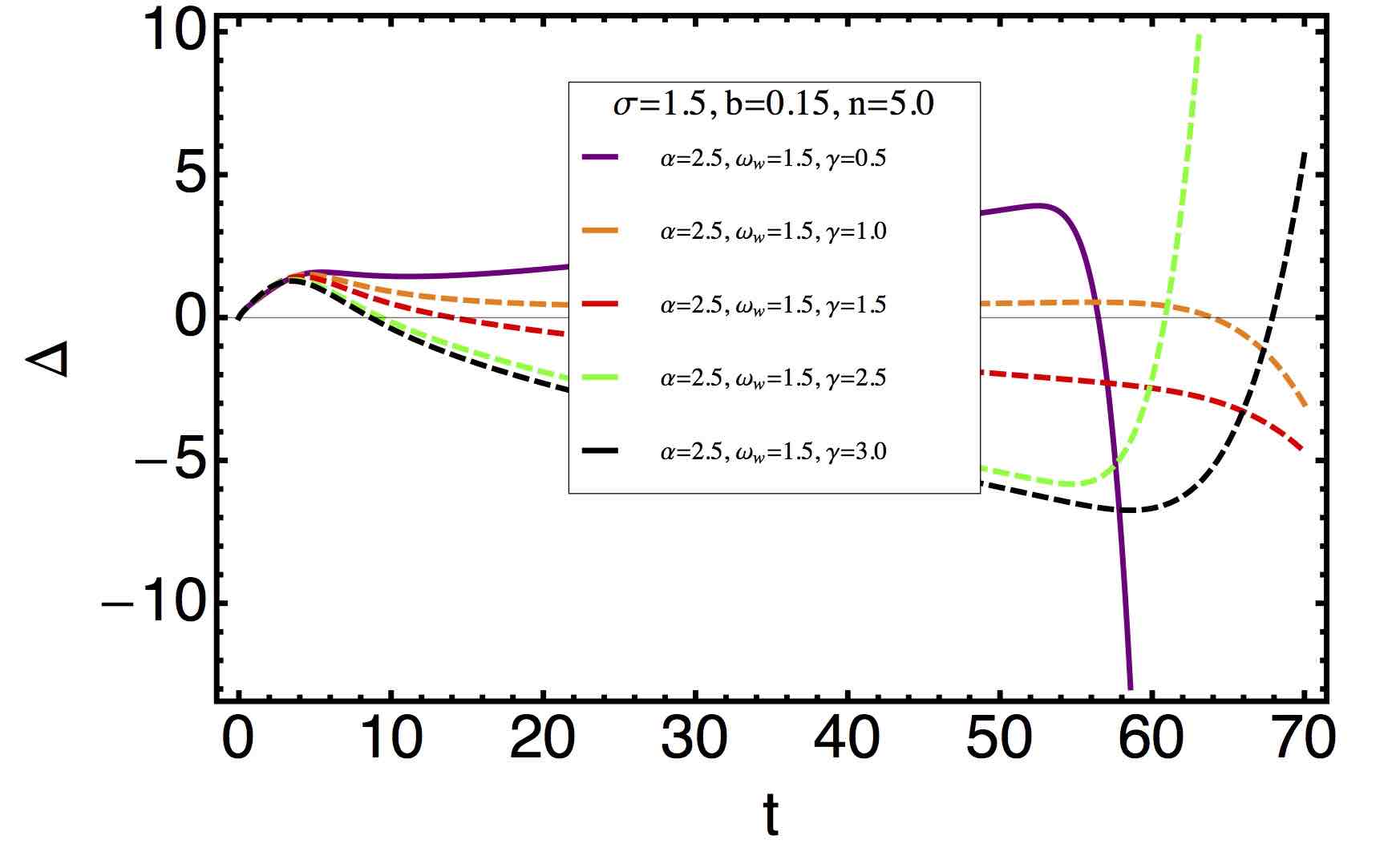}\\
\includegraphics[width=50 mm]{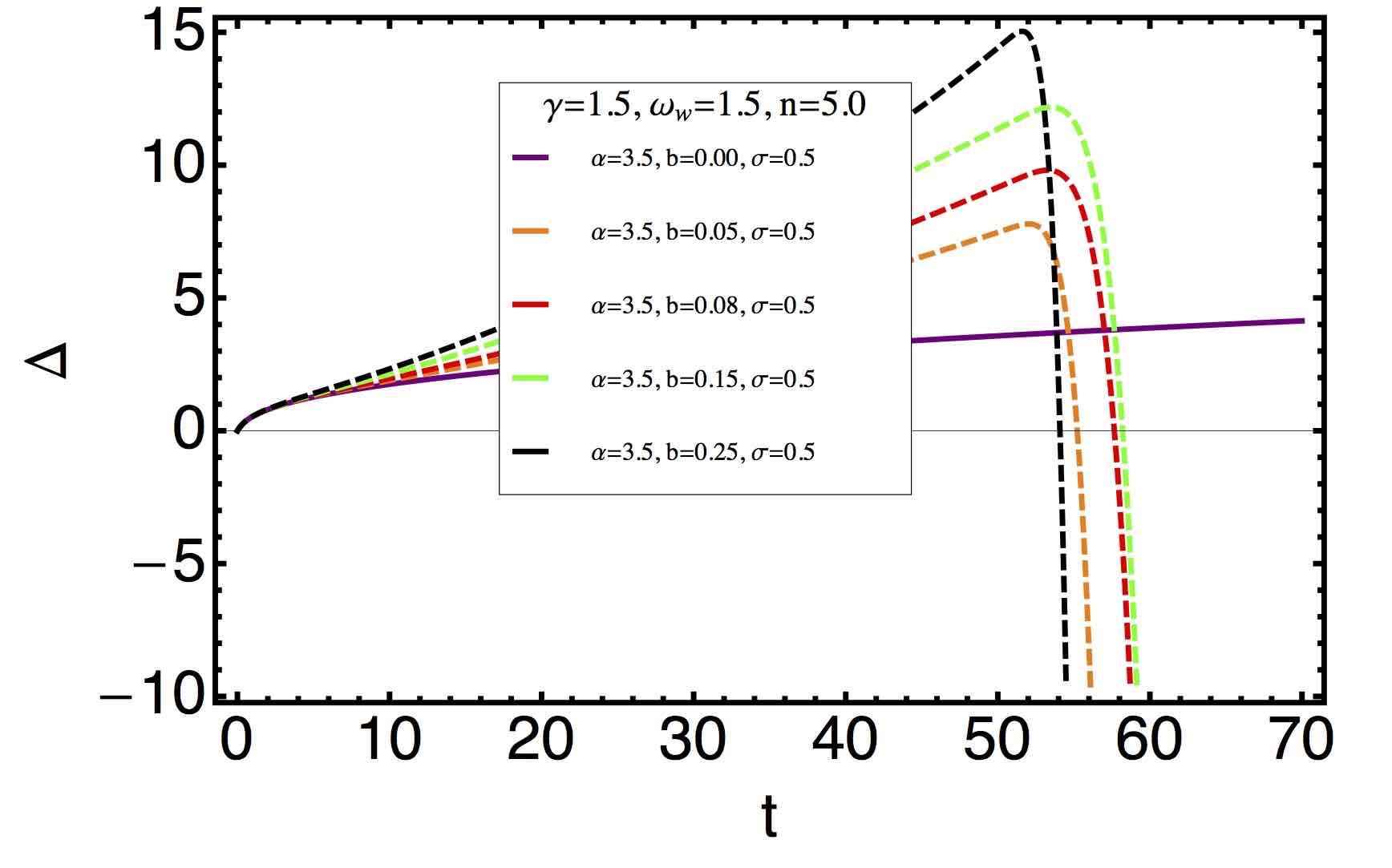} &
\includegraphics[width=50 mm]{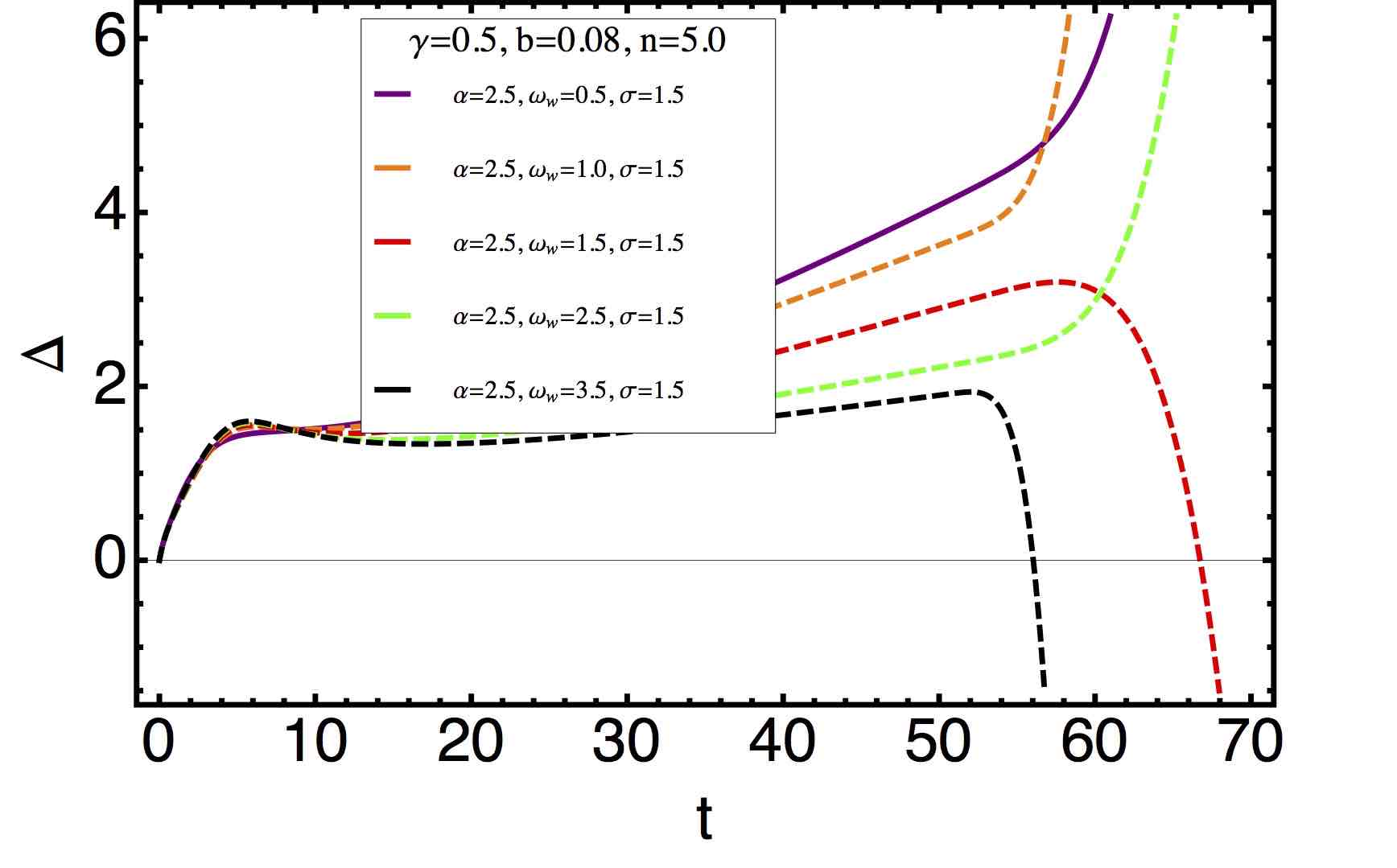}
 \end{array}$
 \end{center}
\caption{Behavior of $\Delta$ against $t$. $\Delta(0)=0$ and $\dot{\Delta}(0)=1$.}
 \label{fig:1}
\end{figure}
\begin{figure}[h!]
 \begin{center}$
 \begin{array}{cccc}
\includegraphics[width=50 mm]{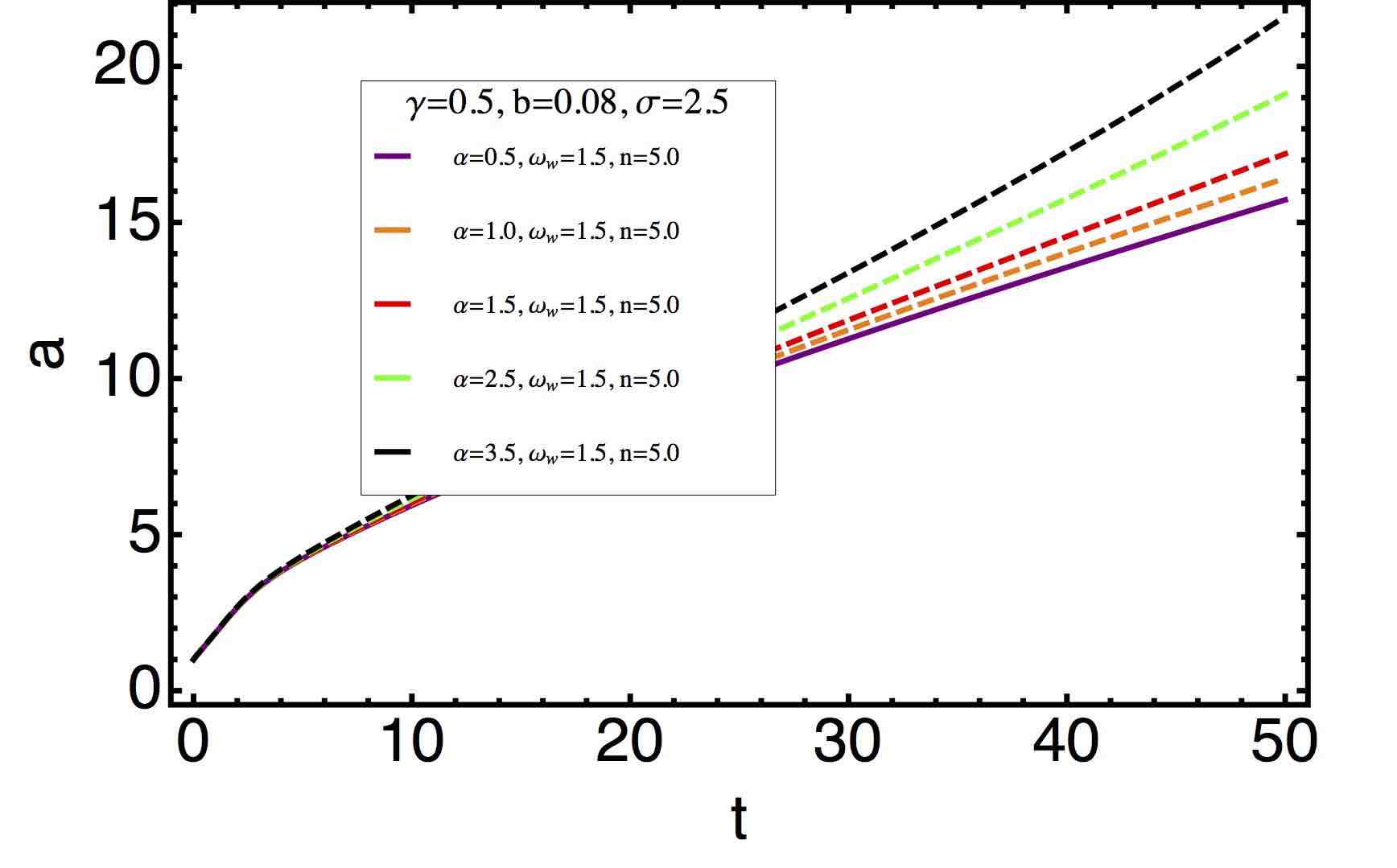} &
\includegraphics[width=50 mm]{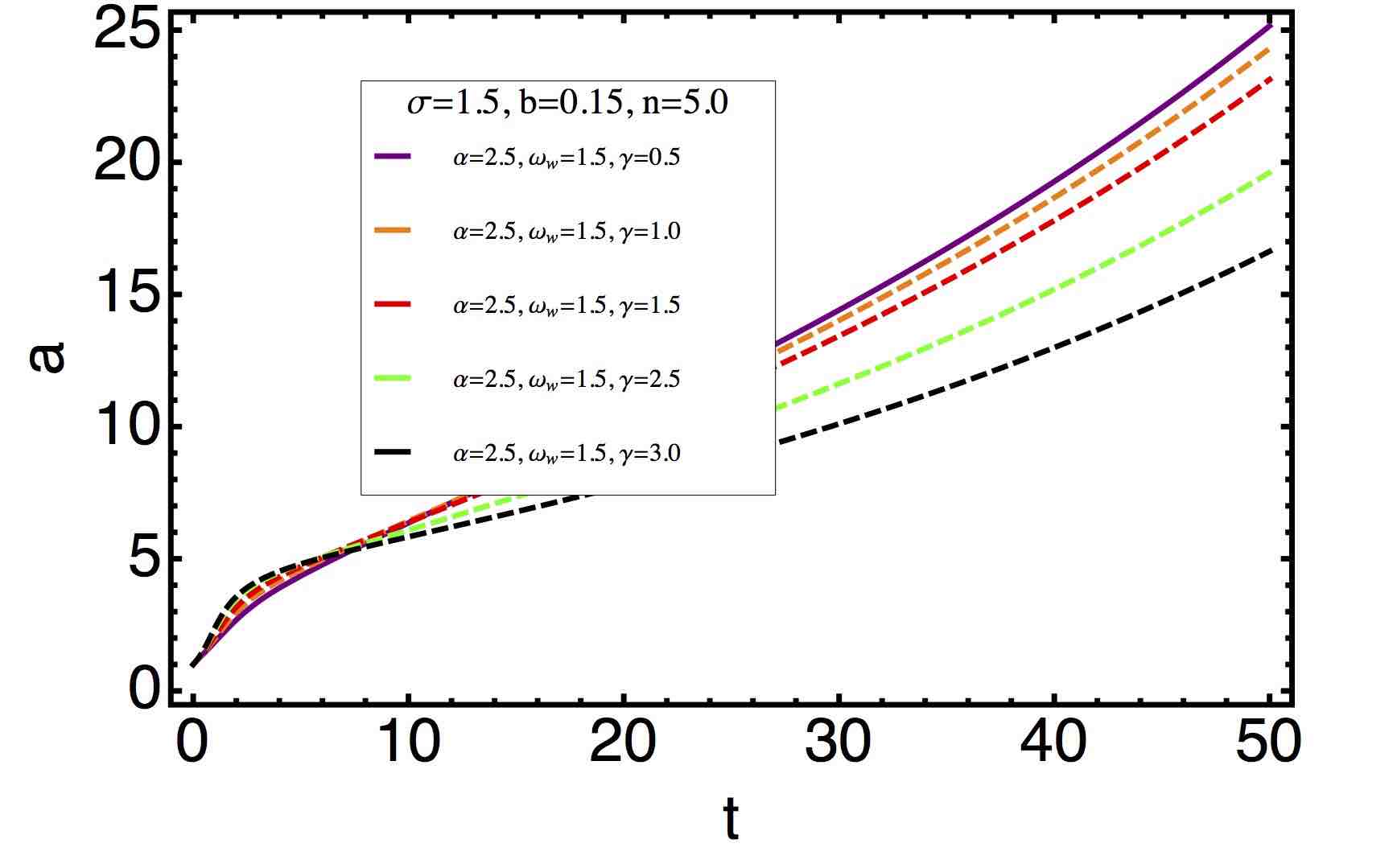}\\
\includegraphics[width=50 mm]{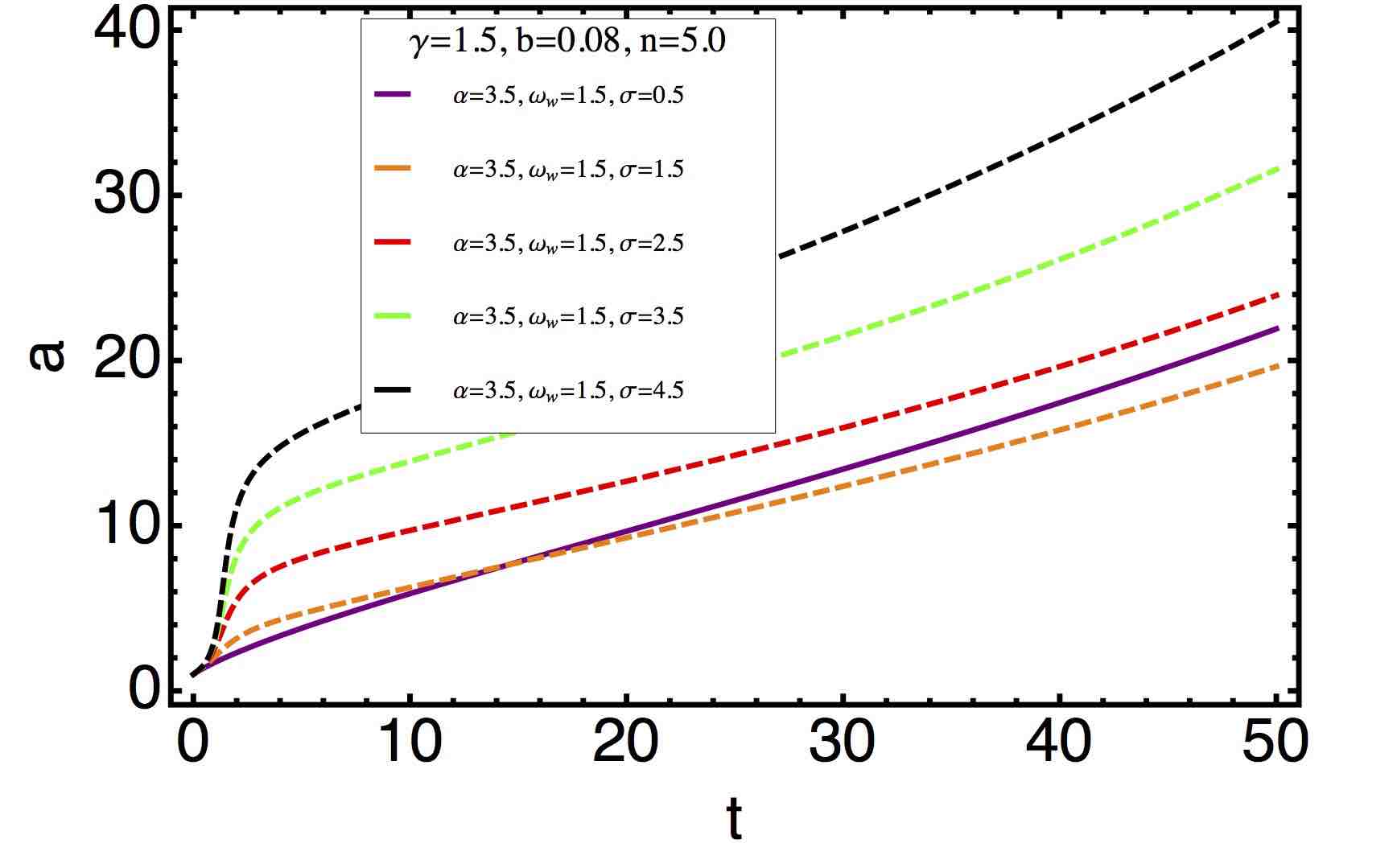} &
\includegraphics[width=50 mm]{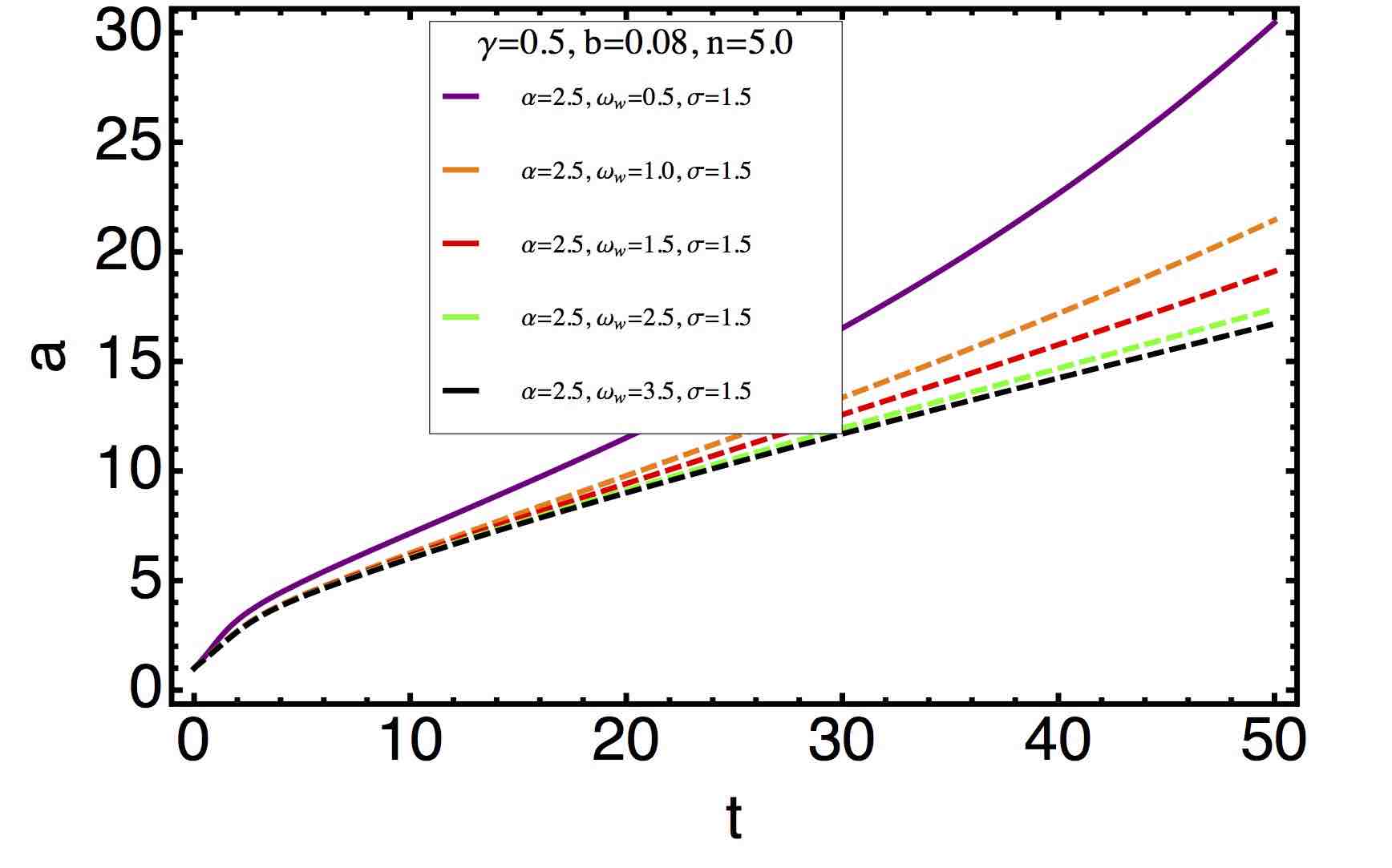}
 \end{array}$
 \end{center}
\caption{Behavior of scale factor $a$ against $t$.}
 \label{fig:1scale}
\end{figure}

However, the speed of sound is not enough to verity stability of system. There are several ways to investigate stability of a theory such as evolution of density perturbations which yields to the following differential equation
\begin{equation}
\ddot{\Delta}+2H\dot{\Delta}-(\frac{\rho}{2}-\frac{C^{2}_{s}}{a^{2}})\Delta=0,
\end{equation}
where $\Delta=\delta \rho / \rho$. Behavior of $\Delta$ over time $t$ as a function of different parameters of the model is presented in Fig. \ref{fig:1}. Analysis of this section were obtained for a scale factor, which profile behavior over time as a function of models parameters can be found in Fig. \ref{fig:1scale}.

\section{\large{Cosmological parameters and numerical results of the case ghost dark energy}}
We solve equations numerically and obtained behavior of $\omega_{tot}$, $P_{tot}$, $\rho_{tot}$. Range of numerical values of the model parameters were fixed based on experimental data as well as to satisfy second law of thermodynamics discussed in Appendix.\\
Plots of Fig. 4 show evolution of total EoS which yields to the -1 as expected as well as behavior of the deceleration parameter (see Fig. 5). However at the initial stages we can see periodic-like behavior. Also we can see acceleration to deceleration phase transition. For the special choice of interaction parameter the periodic like behavior vanishes which is illustrated in the plots of Fig. 6.

\begin{figure}[h!]
 \begin{center}$
 \begin{array}{cccc}
\includegraphics[width=50 mm]{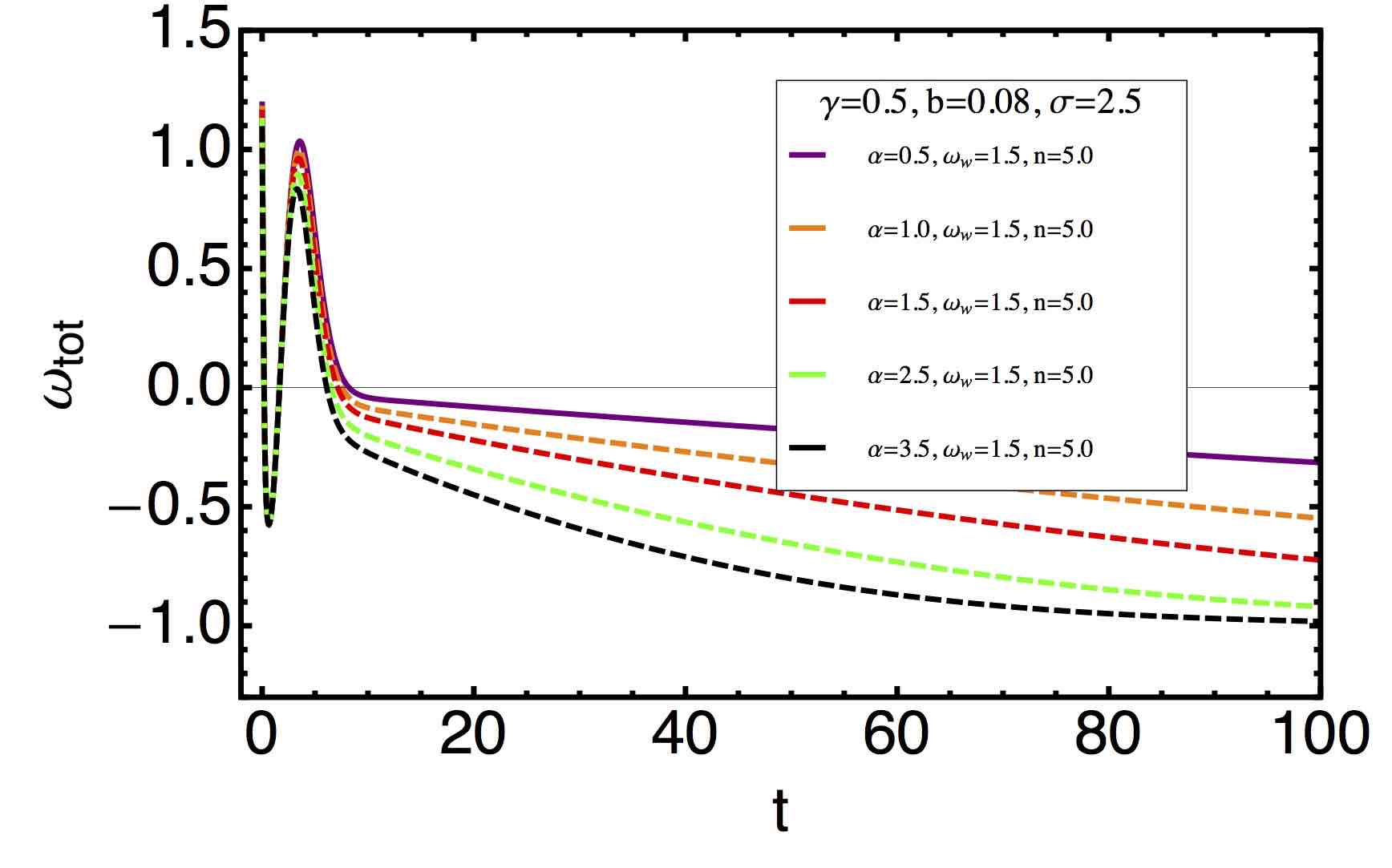} &
\includegraphics[width=50 mm]{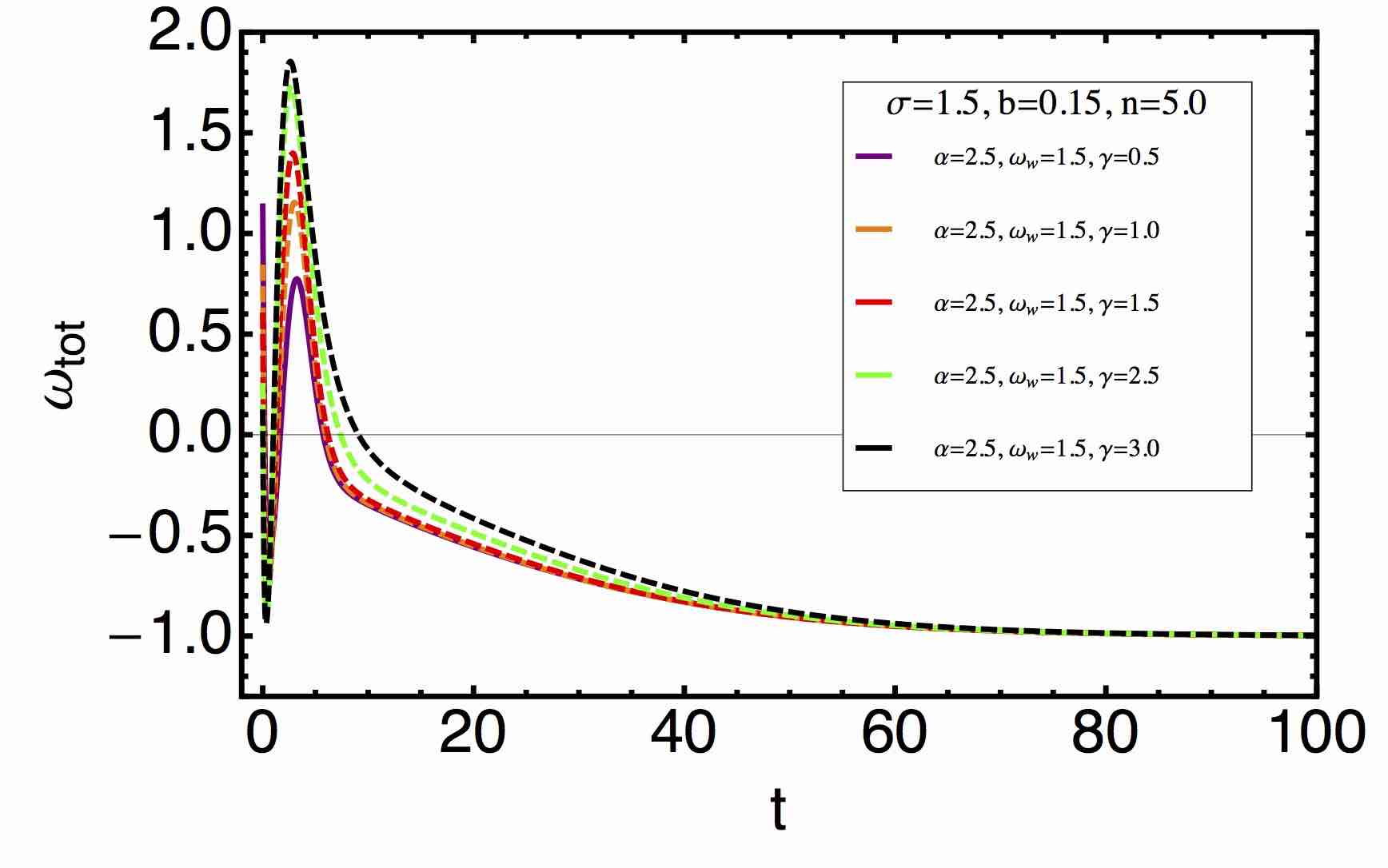}\\
\includegraphics[width=50 mm]{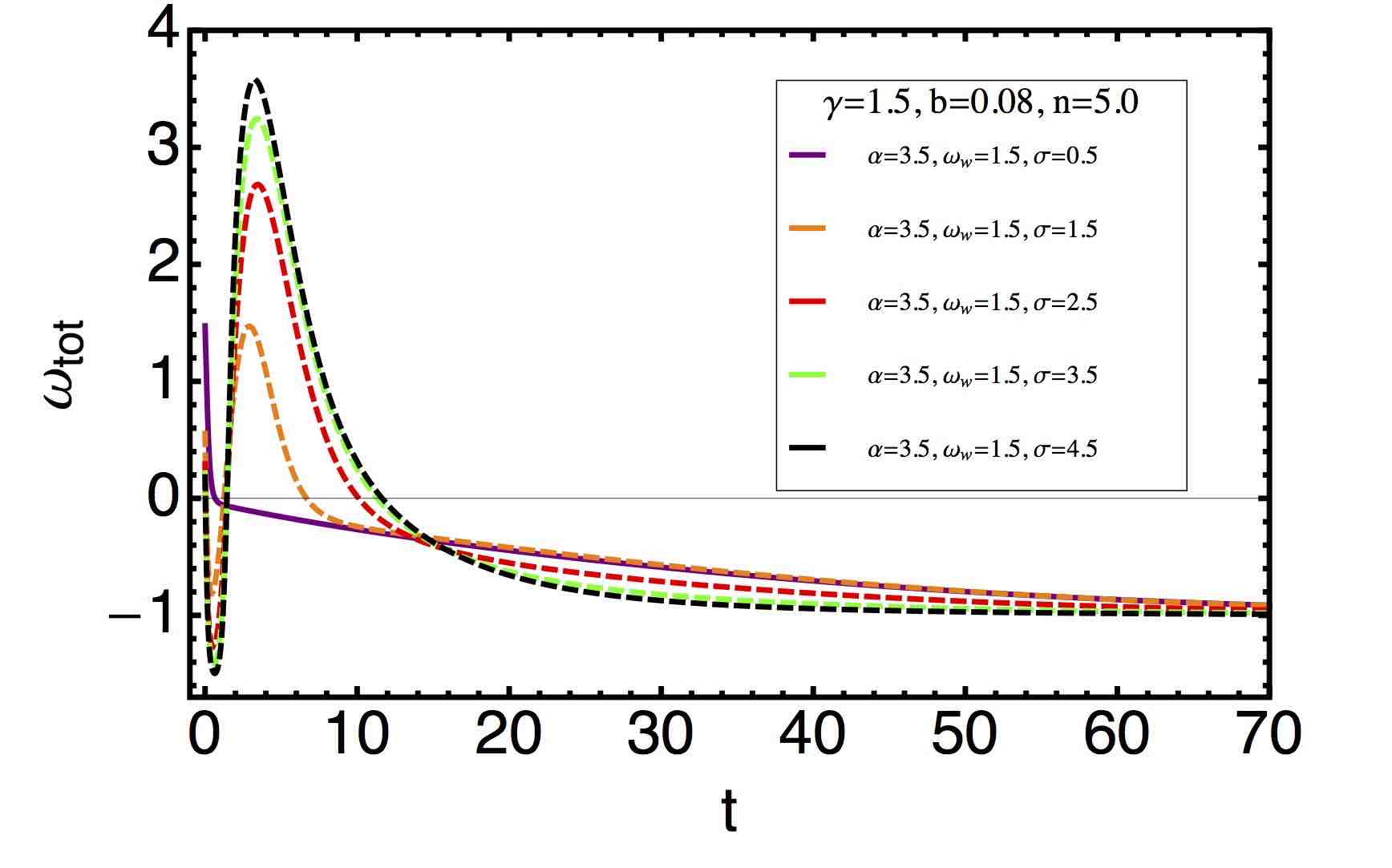} &
\includegraphics[width=50 mm]{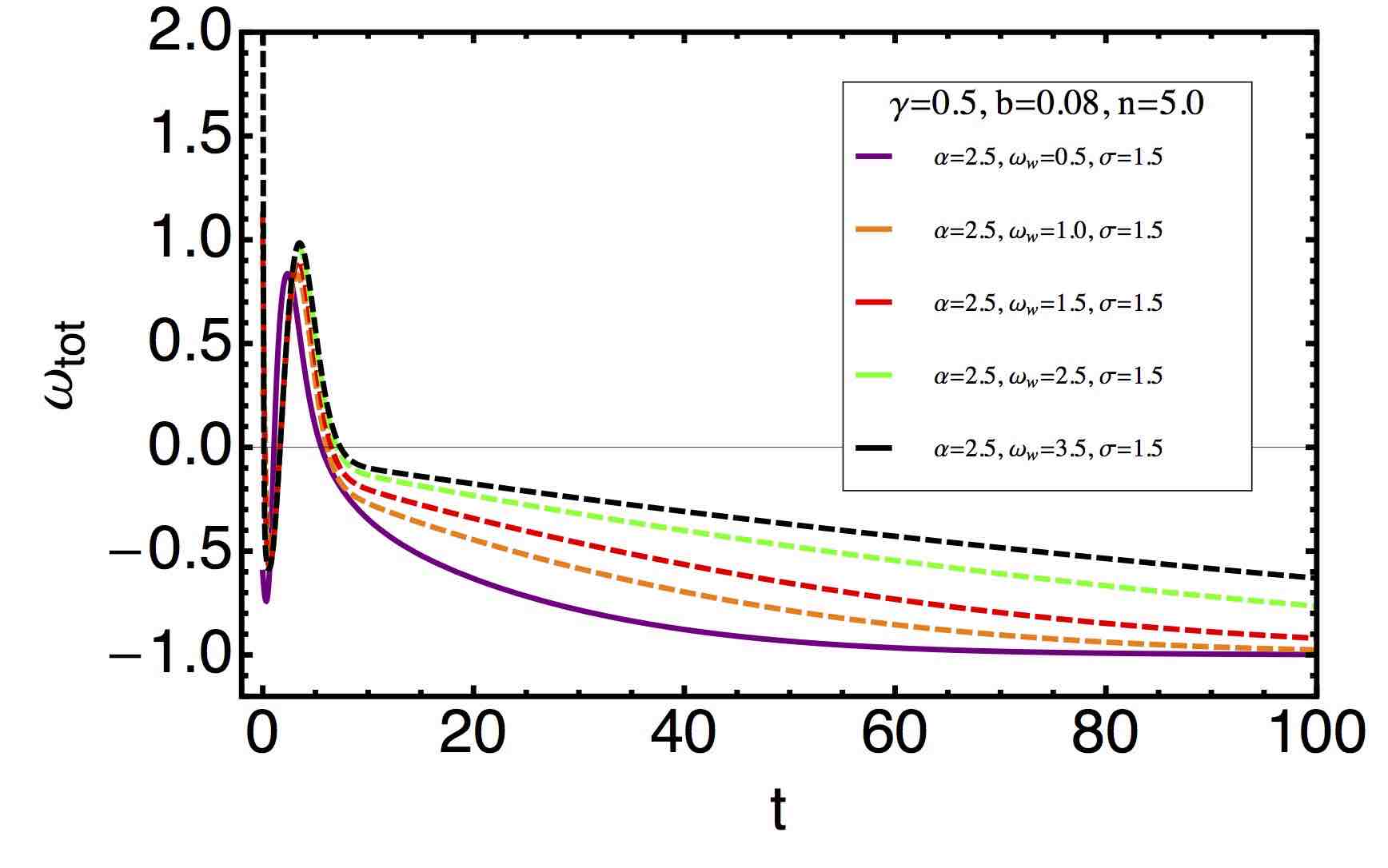}
 \end{array}$
 \end{center}
\caption{Behavior of scale factor $\omega_{tot}$ against $t$.}
 \label{fig:2}
\end{figure}

\begin{figure}[h!]
 \begin{center}$
 \begin{array}{cccc}
\includegraphics[width=50 mm]{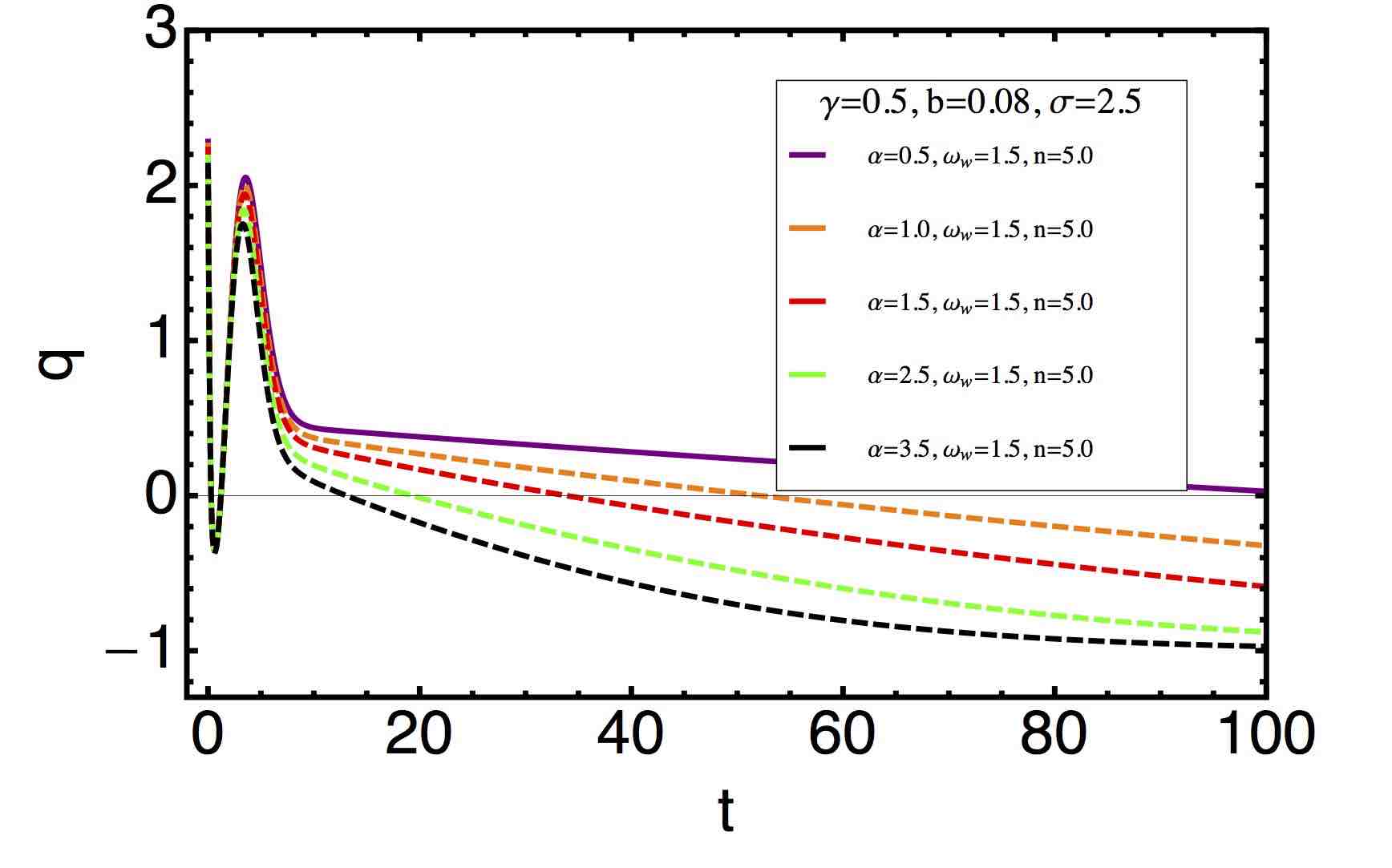} &
\includegraphics[width=50 mm]{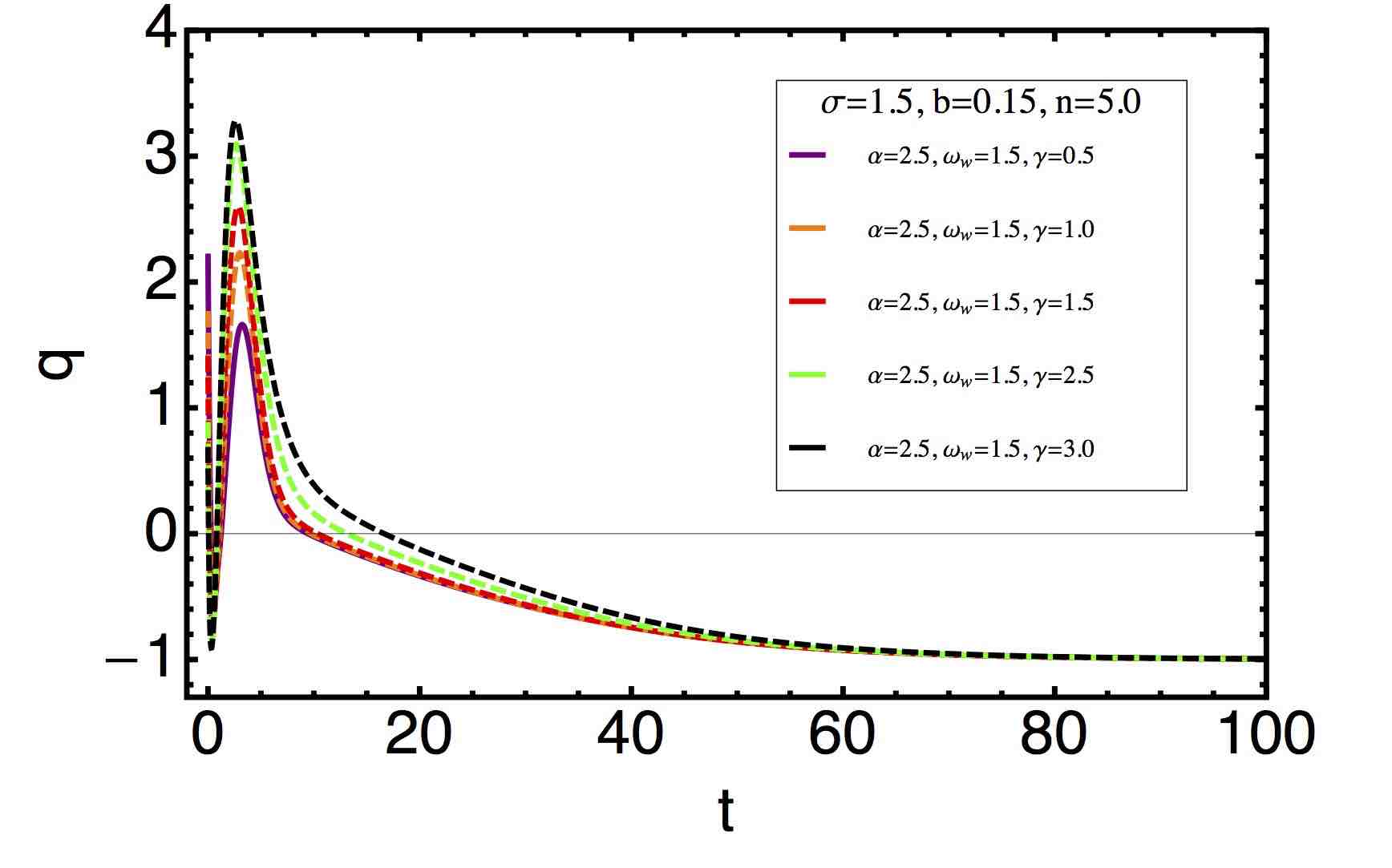}\\
\includegraphics[width=50 mm]{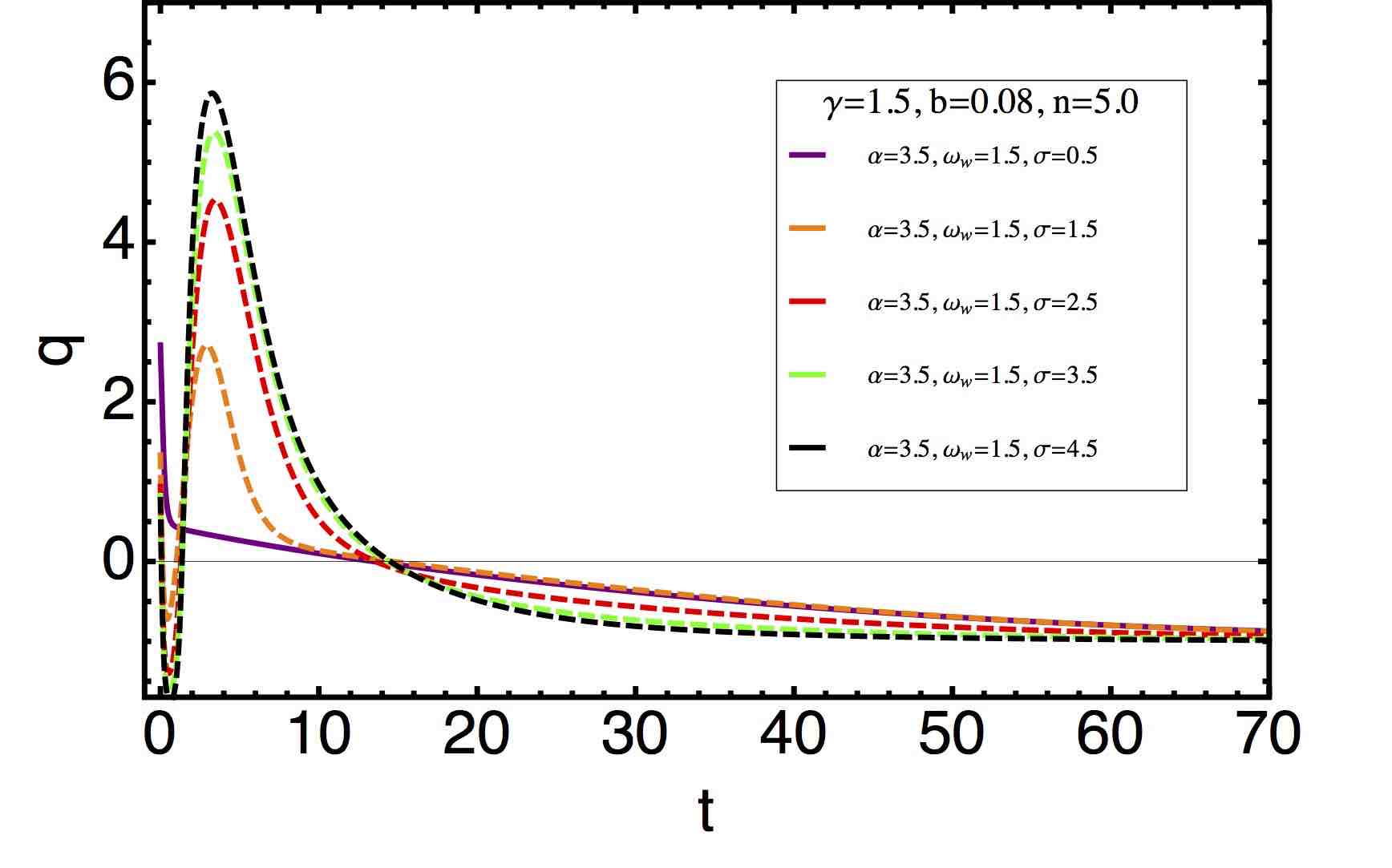} &
\includegraphics[width=50 mm]{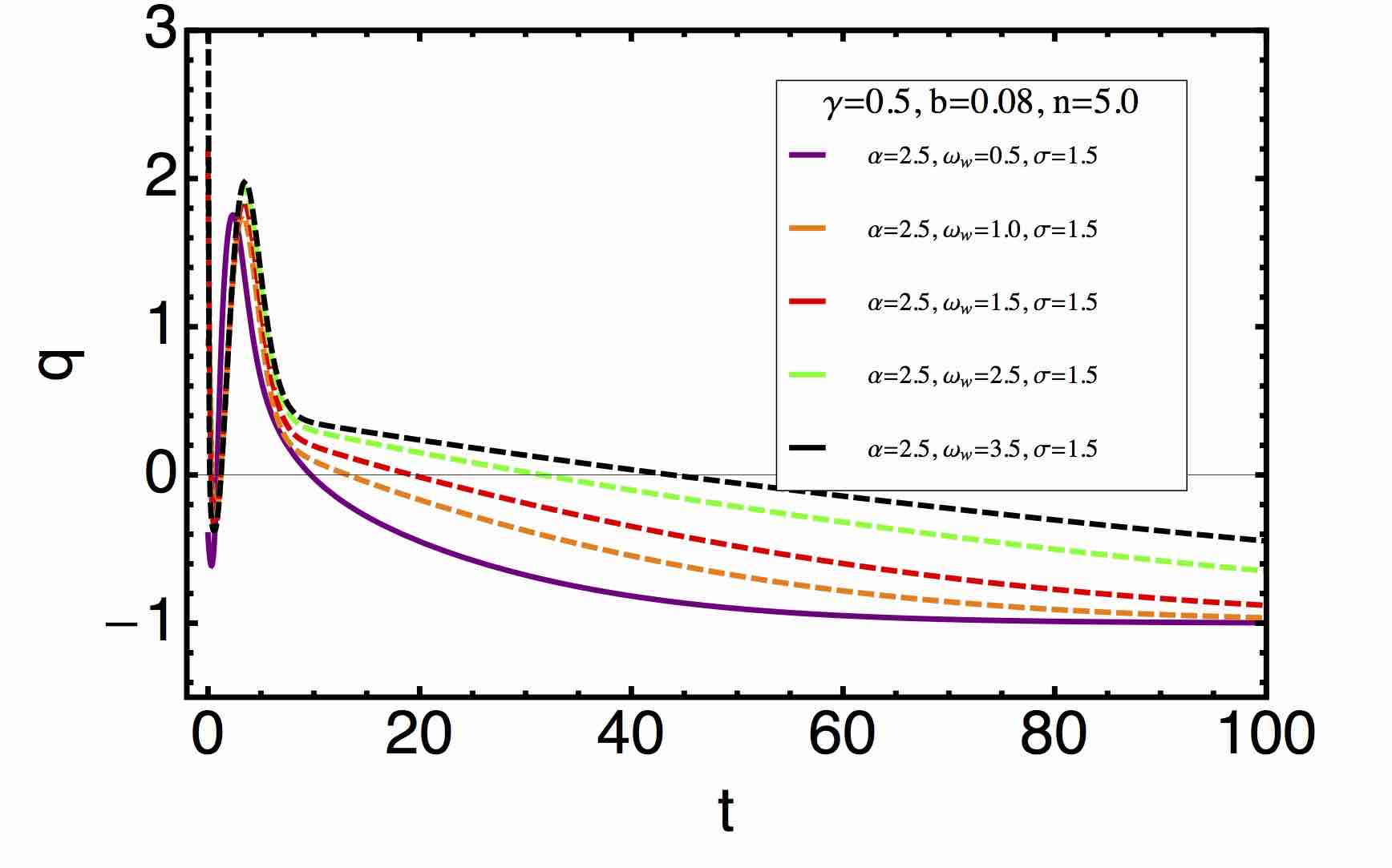}
 \end{array}$
 \end{center}
\caption{Behavior of deceleration parameter $q$ against $t$.}
 \label{fig:3}
\end{figure}

\begin{figure}[h!]
 \begin{center}$
 \begin{array}{cccc}
\includegraphics[width=50 mm]{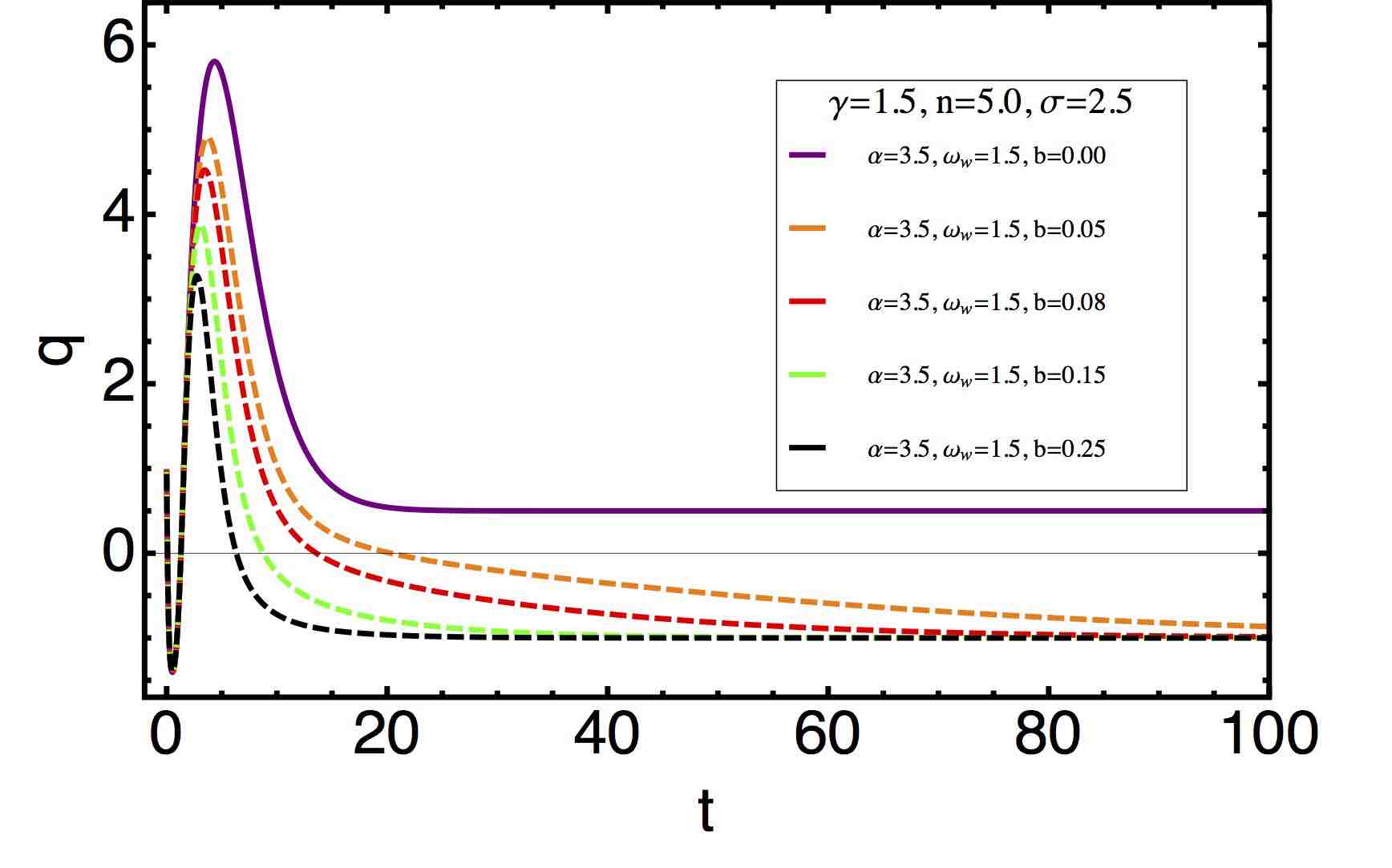} &
\includegraphics[width=50 mm]{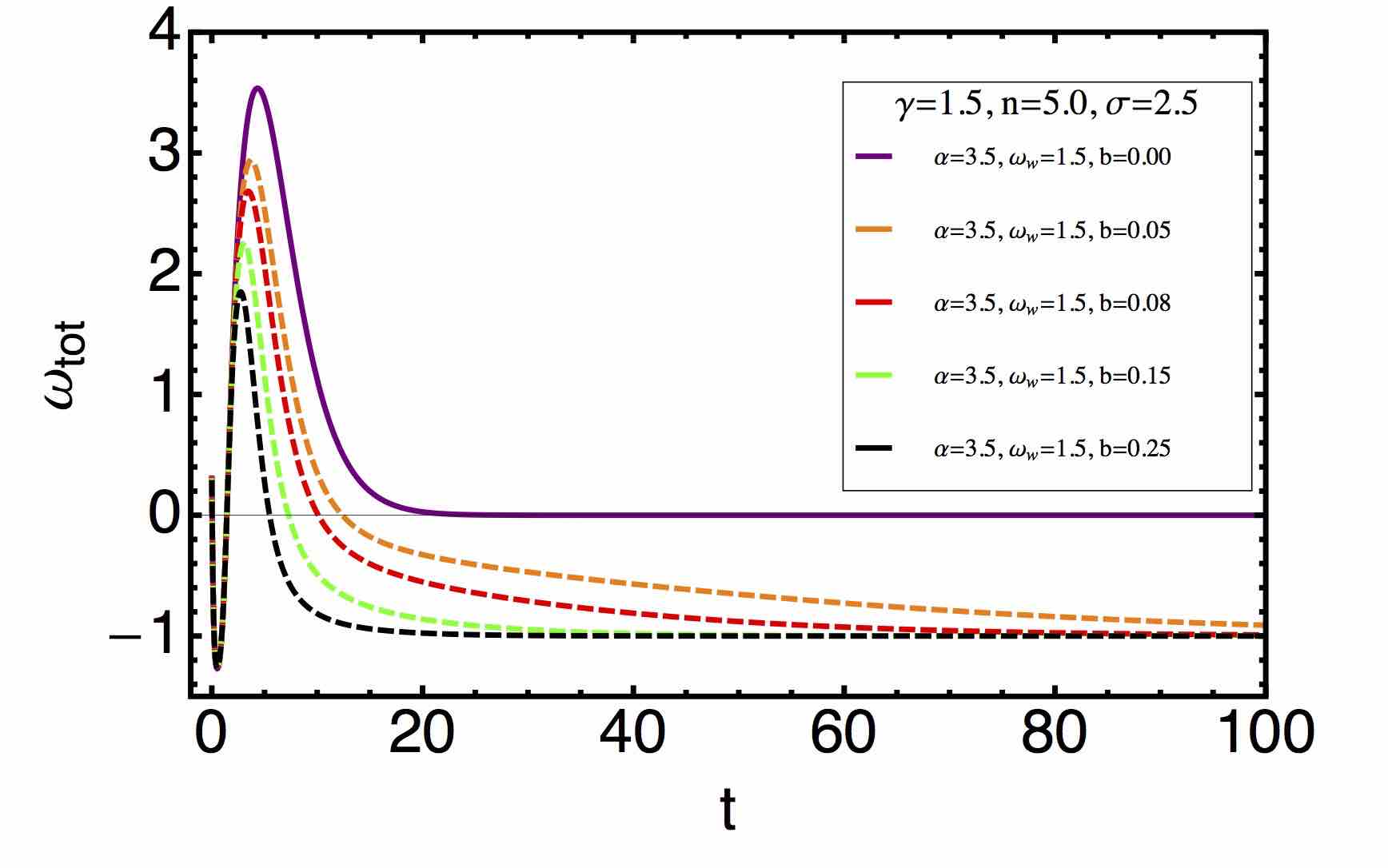}\\
\includegraphics[width=50 mm]{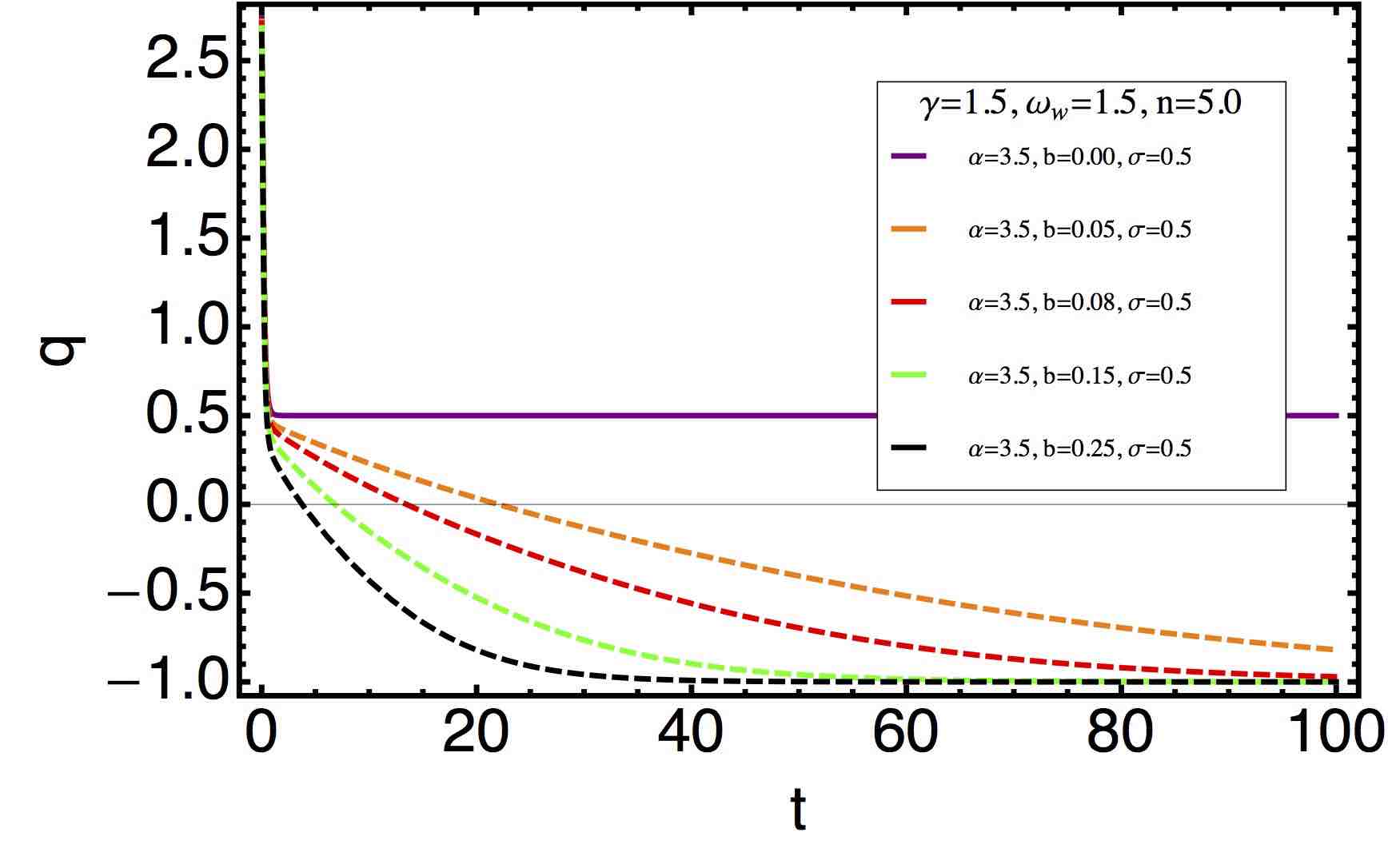} &
\includegraphics[width=50 mm]{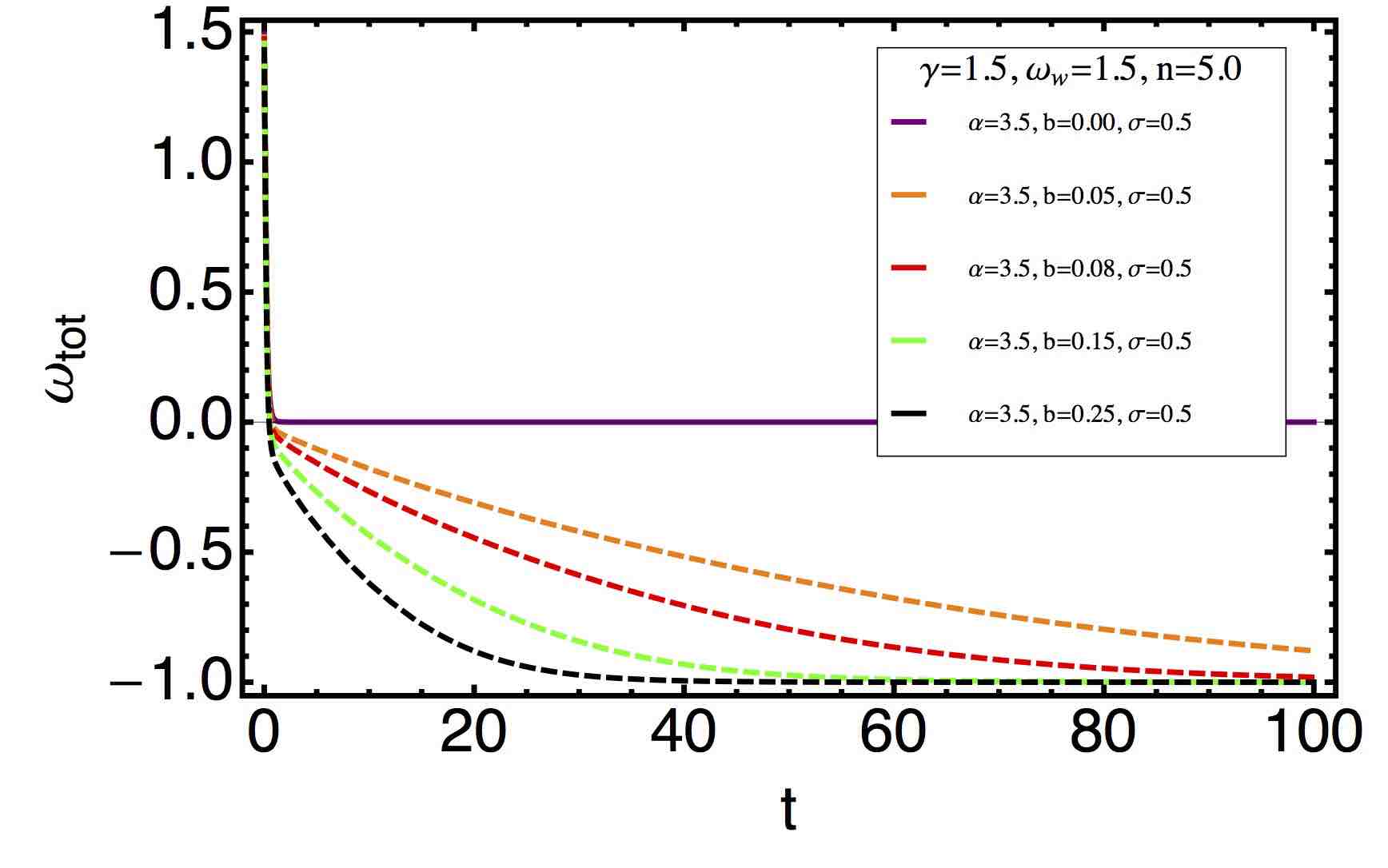}
 \end{array}$
 \end{center}
\caption{Behavior of deceleration parameter $q$ against $t$ for different values of interaction parameter $b$.}
 \label{fig:4}
\end{figure}

\section{\large{Cosmological parameters and numerical results of the case generalized ghost dark energy}}
In this section we consider the case of generalized ghost dark energy and discuss about cosmological parameters and stability of the model numerically.\\
The plots of Fig. 7 show there are some stabilities before late time stages.\\
Also, we can see similar behavior with the previous case for perturbation $\Delta$ (see Fig. 8).\\
Then, in Fig. 9 we can see time evolution of the scale factor and find that it is increasing function as expected. As $\alpha$ and $\sigma$ increased the value of scale factor increases too, but it is decreased by $\omega_{w}$. The variation with $\gamma$ is different at initial and late stage. At the early Universe the value of $a$ grows withy $\gamma$ but at the late stage it is decreased by $\gamma$.\\
Then, we can see the approximately the same behavior for total EoS in Fig. 10 as well as the deceleration parameter in Fig. 11.

\begin{figure}[h!]
 \begin{center}$
 \begin{array}{cccc}
\includegraphics[width=50 mm]{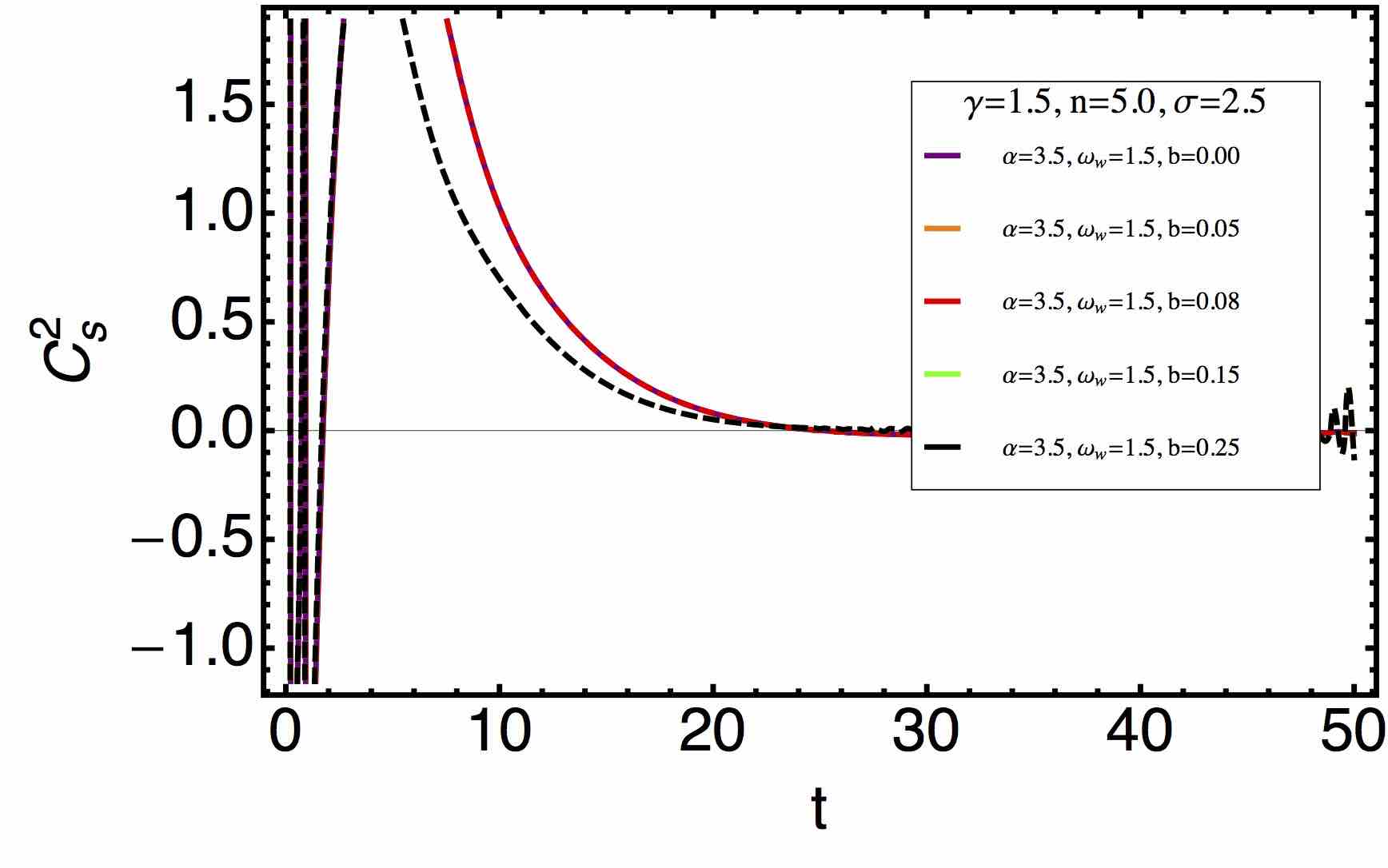} &
\includegraphics[width=50 mm]{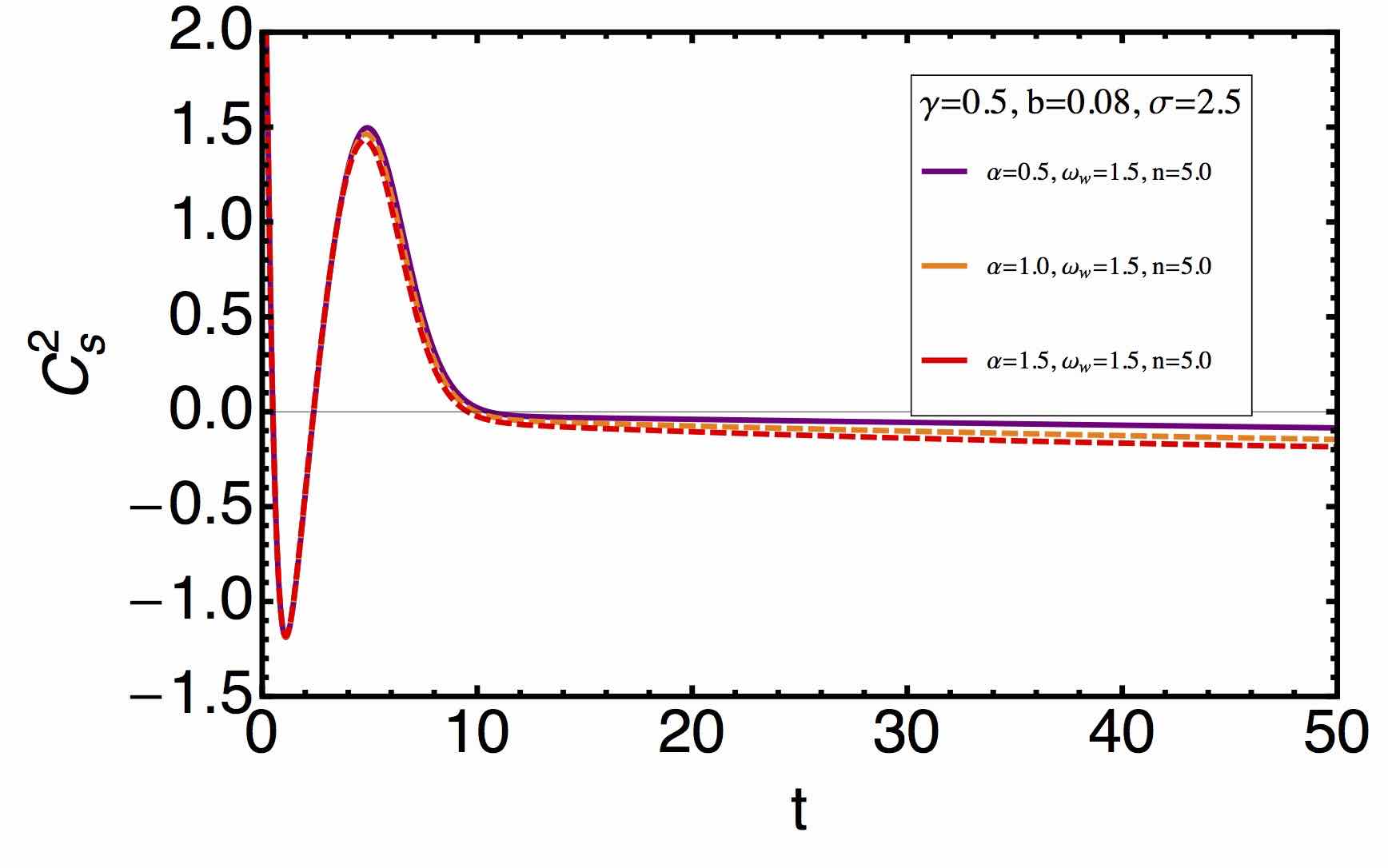}\\
\includegraphics[width=50 mm]{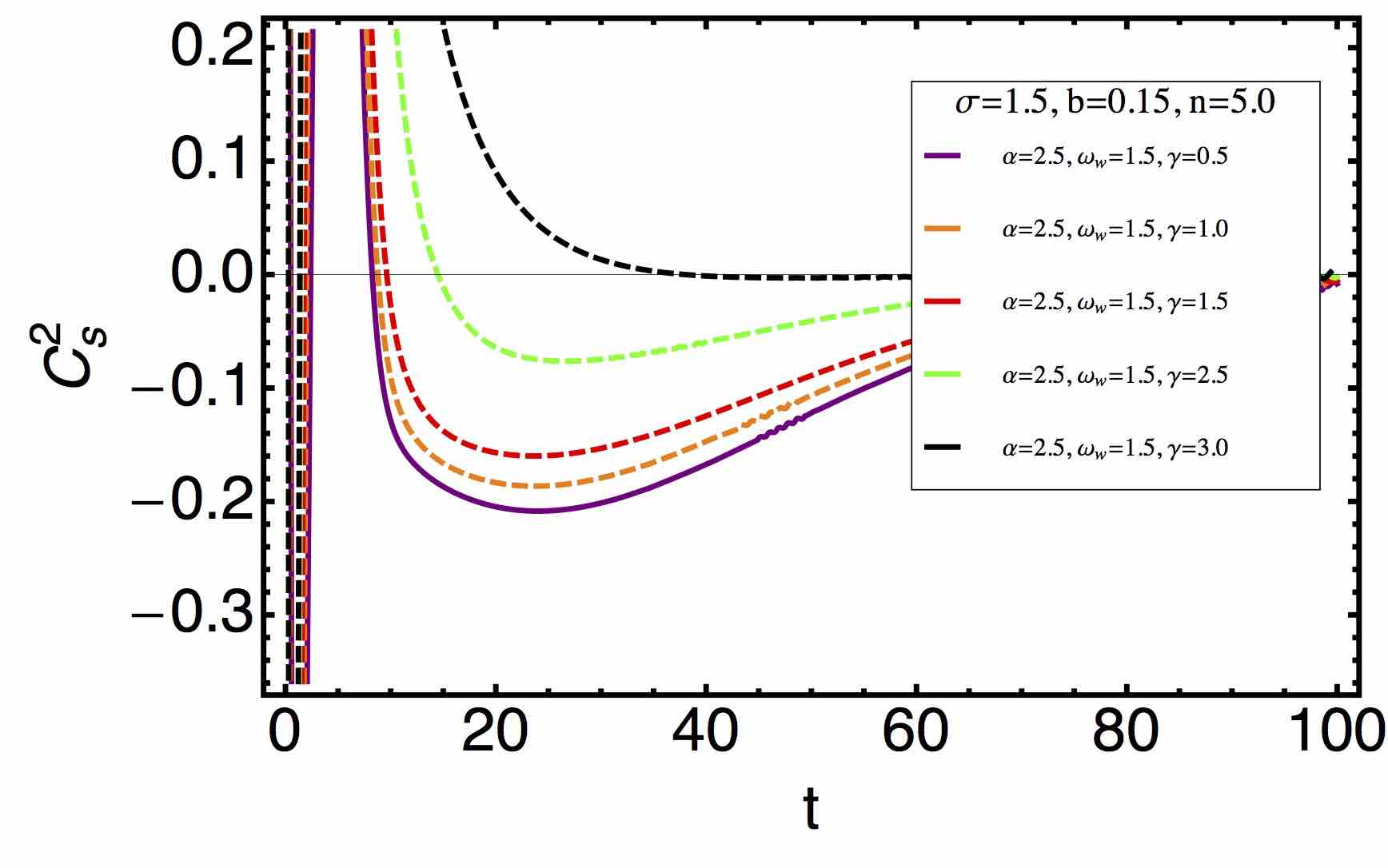} &
\includegraphics[width=50 mm]{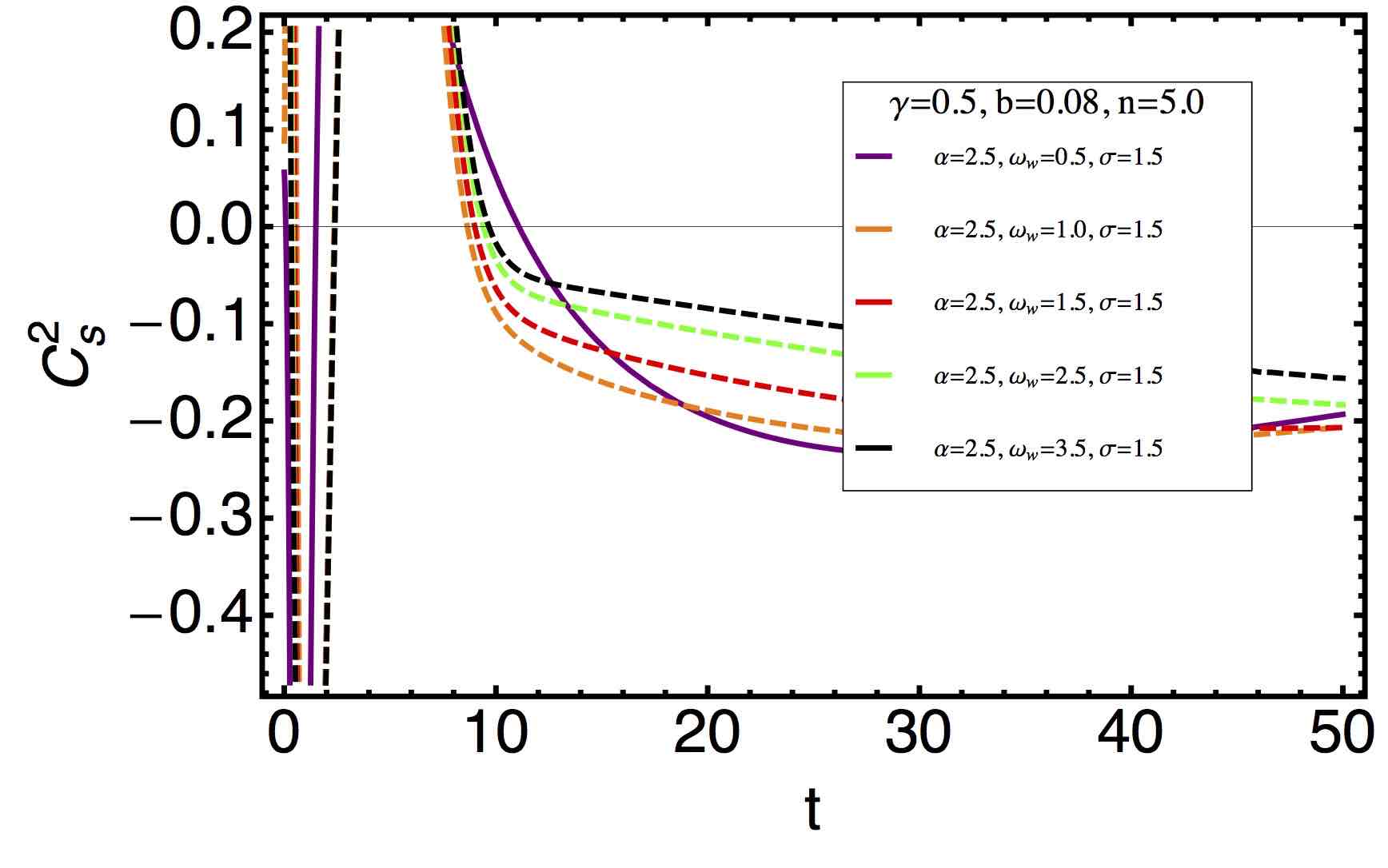}\\
 \end{array}$
 \end{center}
\caption{Behavior of squared sound speed $C_{s}^{2}$ against $t$.}
 \label{fig:7}
\end{figure}

\begin{figure}[h!]
 \begin{center}$
 \begin{array}{cccc}
\includegraphics[width=50 mm]{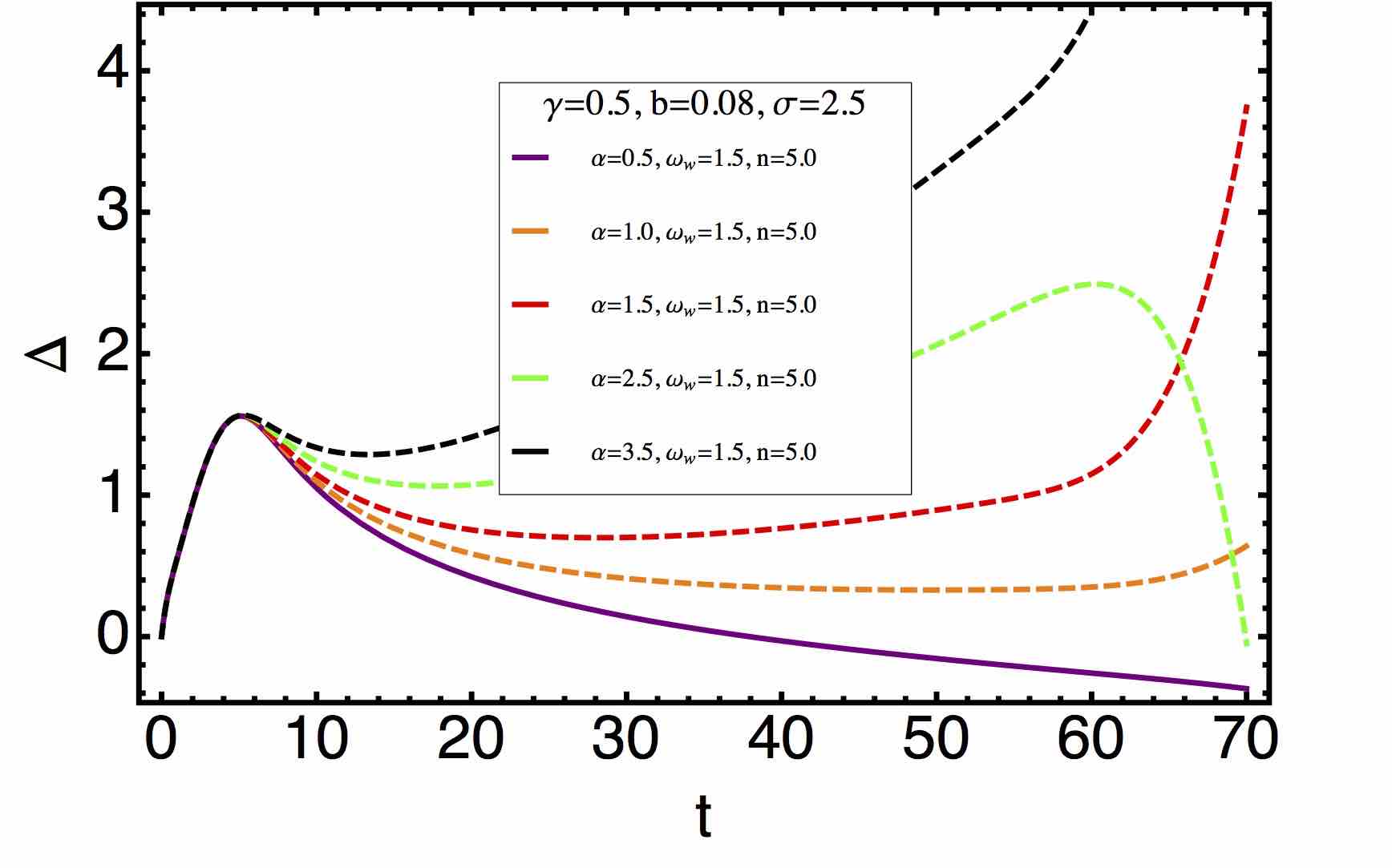} &
\includegraphics[width=50 mm]{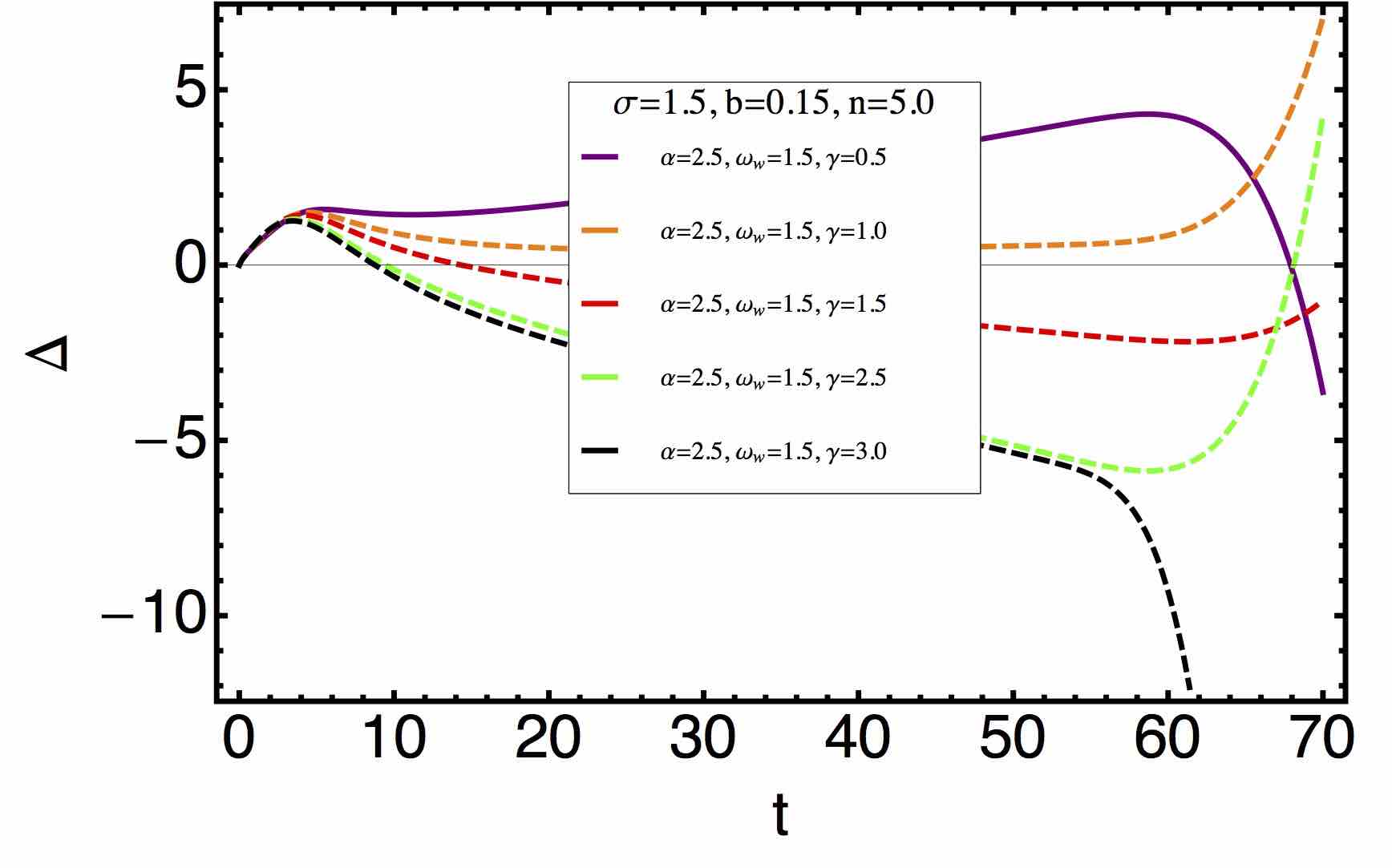}
 \end{array}$
 \end{center}
\caption{Behavior of $\Delta$ against $t$. $\Delta(0)=0$ and $\dot{\Delta}(0)=1$.}
 \label{fig:8}
\end{figure}

\begin{figure}[h!]
 \begin{center}$
 \begin{array}{cccc}
\includegraphics[width=50 mm]{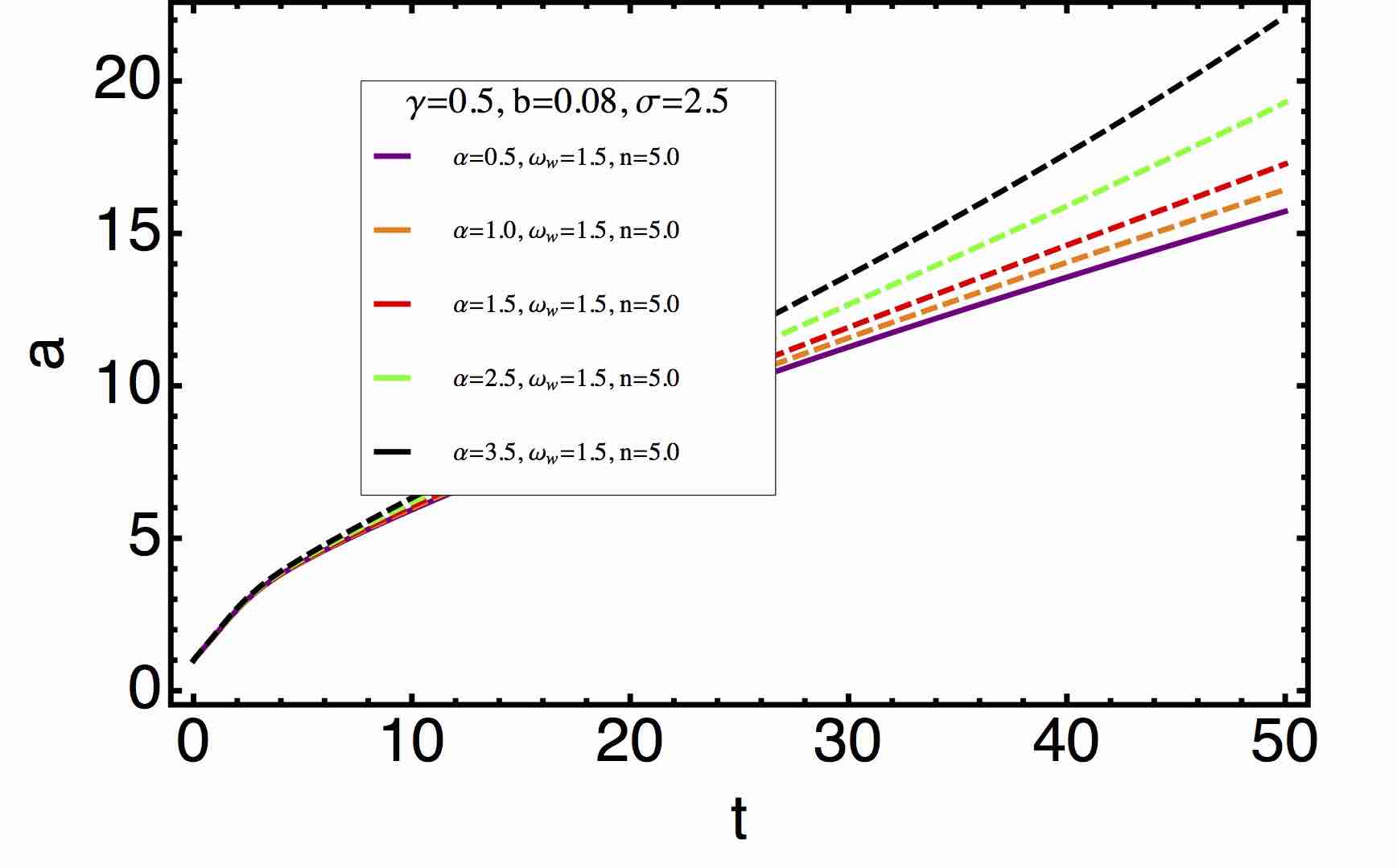} &
\includegraphics[width=50 mm]{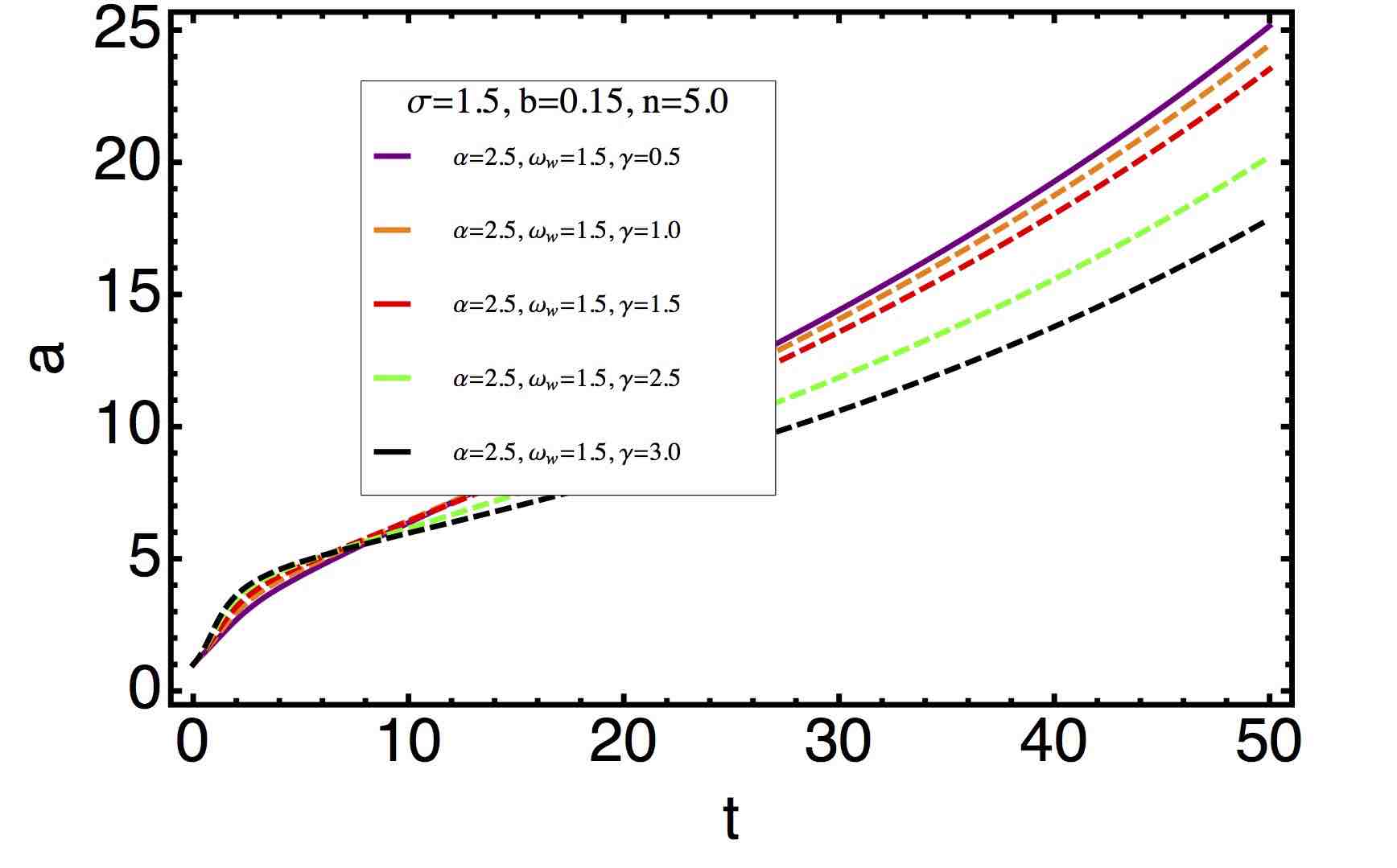}\\
\includegraphics[width=50 mm]{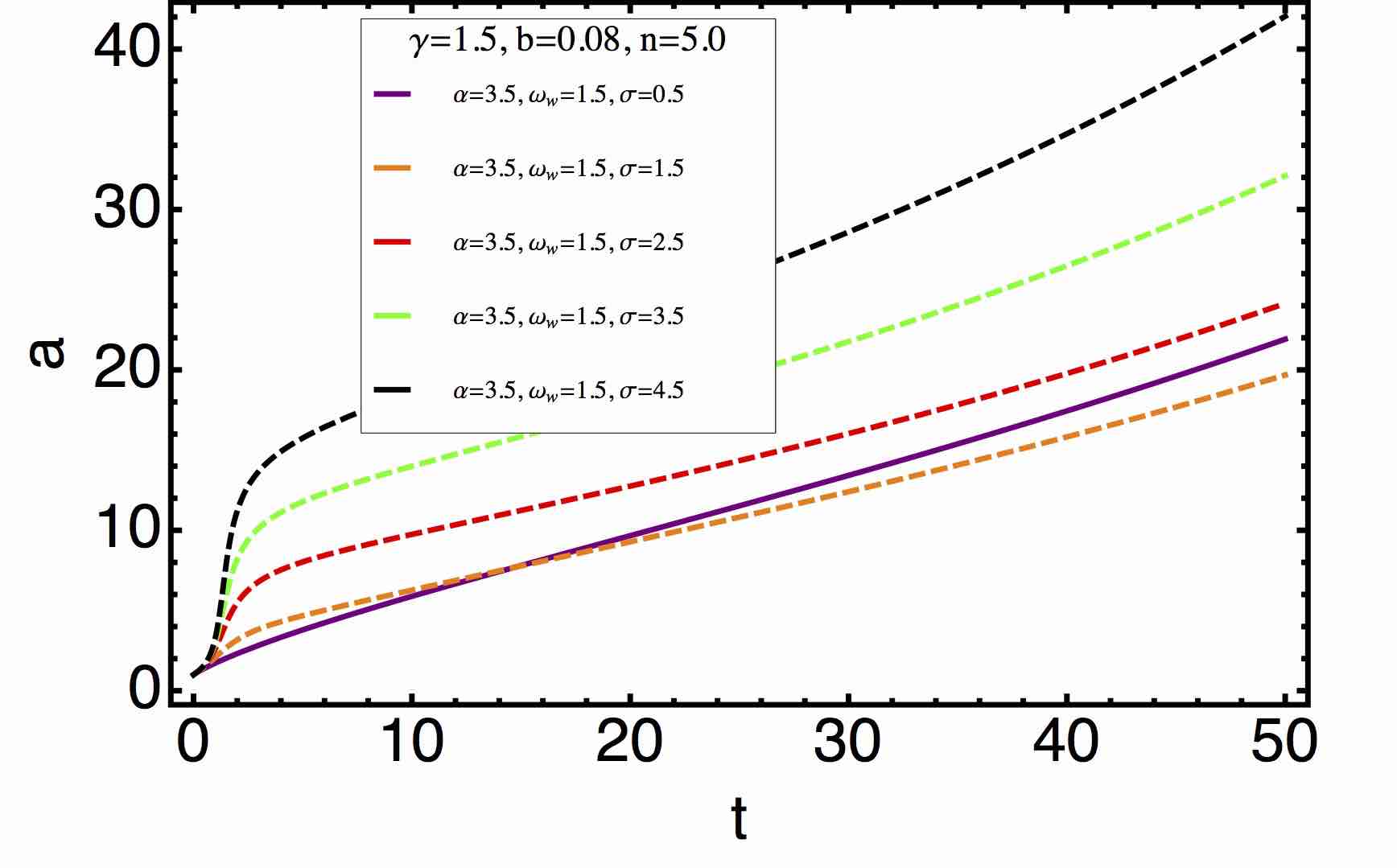} &
\includegraphics[width=50 mm]{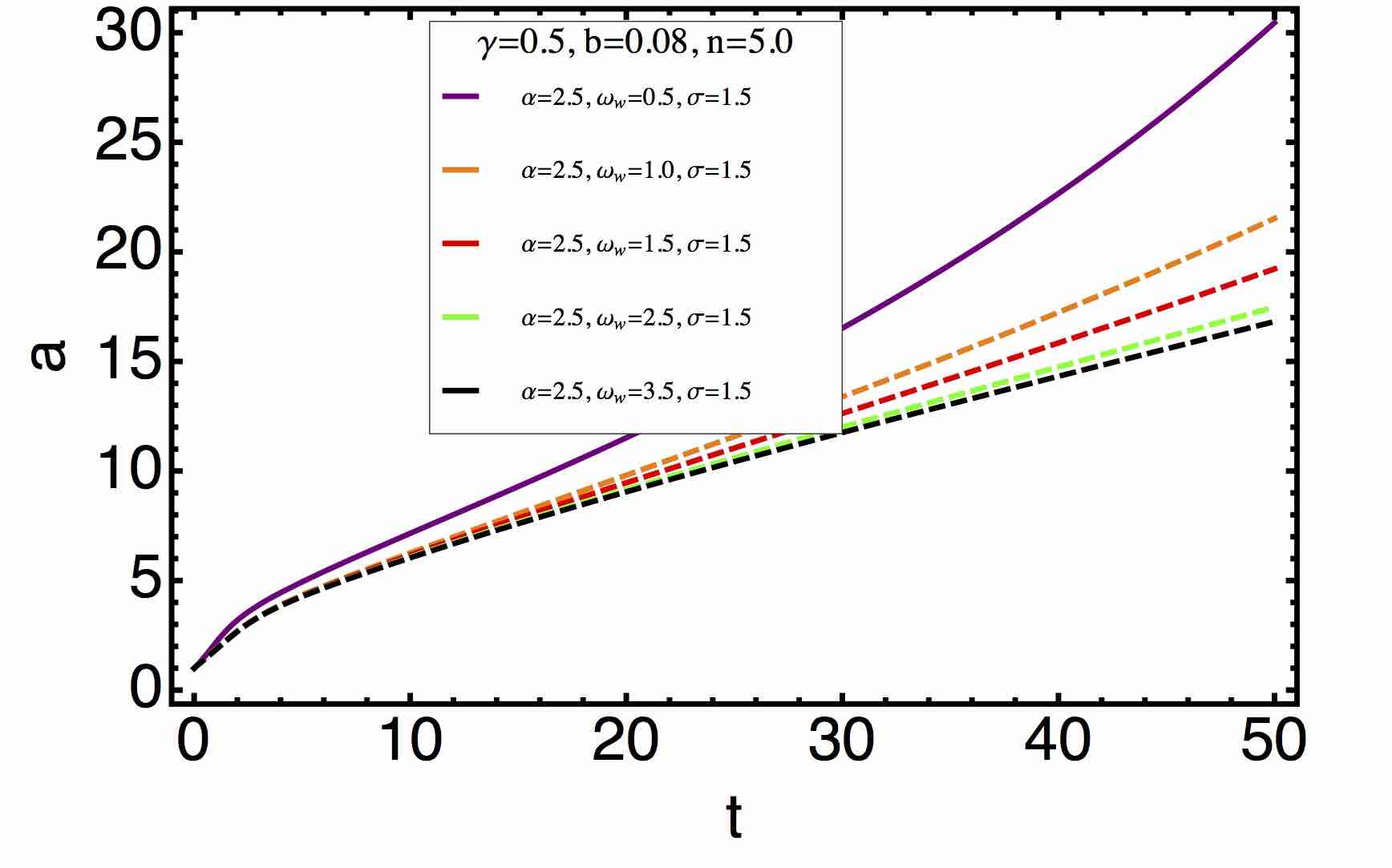}
 \end{array}$
 \end{center}
\caption{Behavior of scale factor $a$ against $t$.}
 \label{fig:9}
\end{figure}

\begin{figure}[h!]
 \begin{center}$
 \begin{array}{cccc}
\includegraphics[width=50 mm]{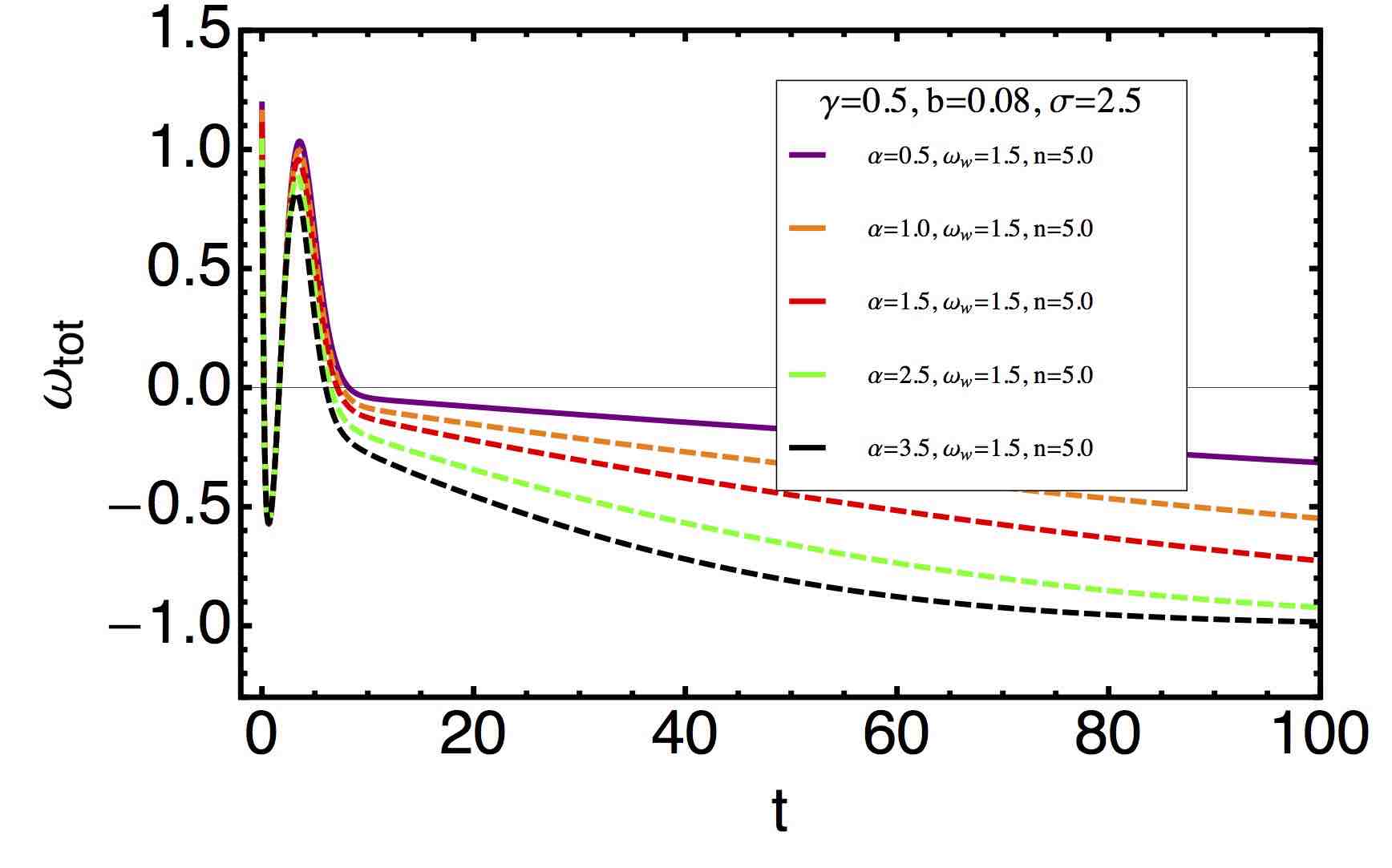} &
\includegraphics[width=50 mm]{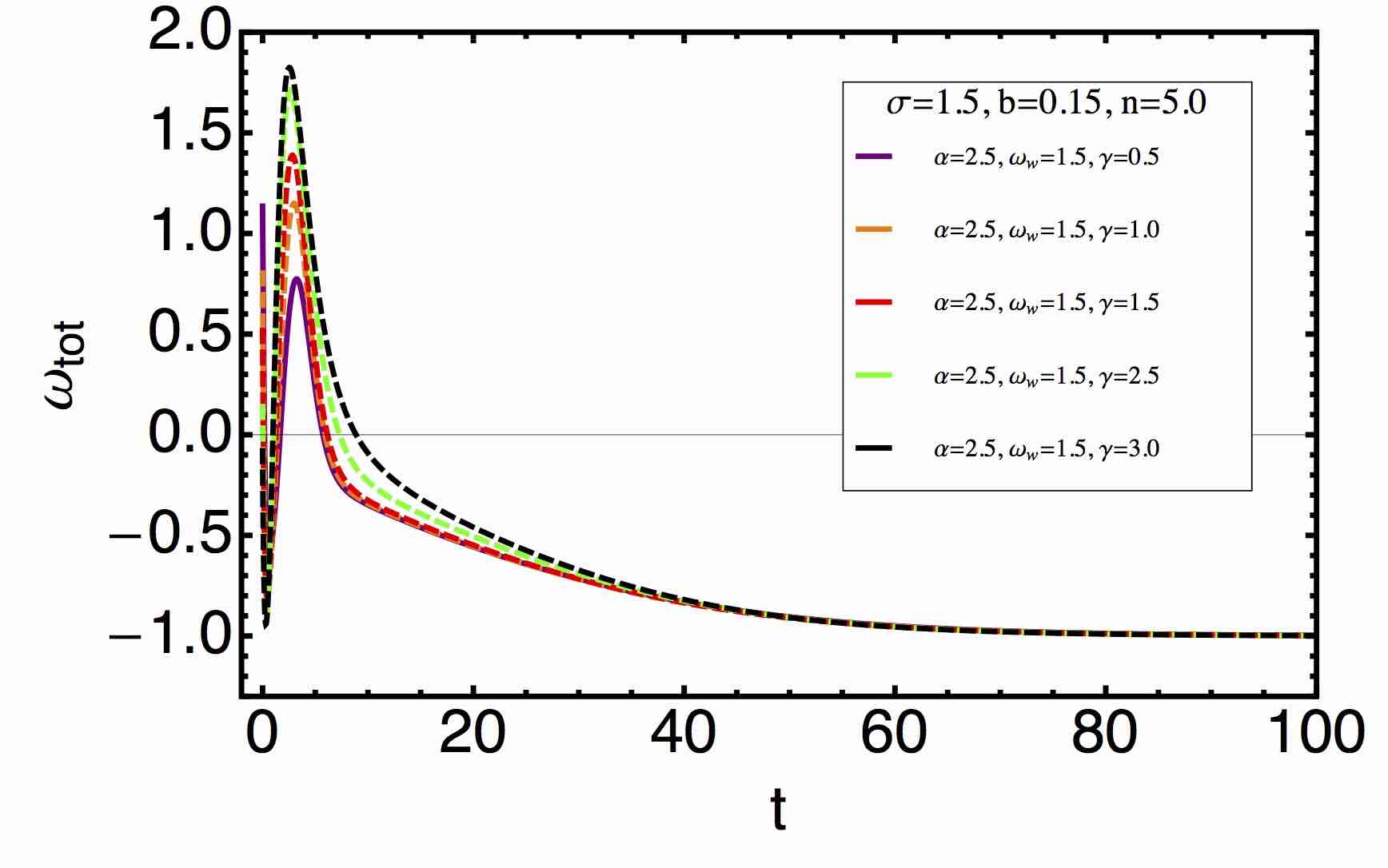}\\
\includegraphics[width=50 mm]{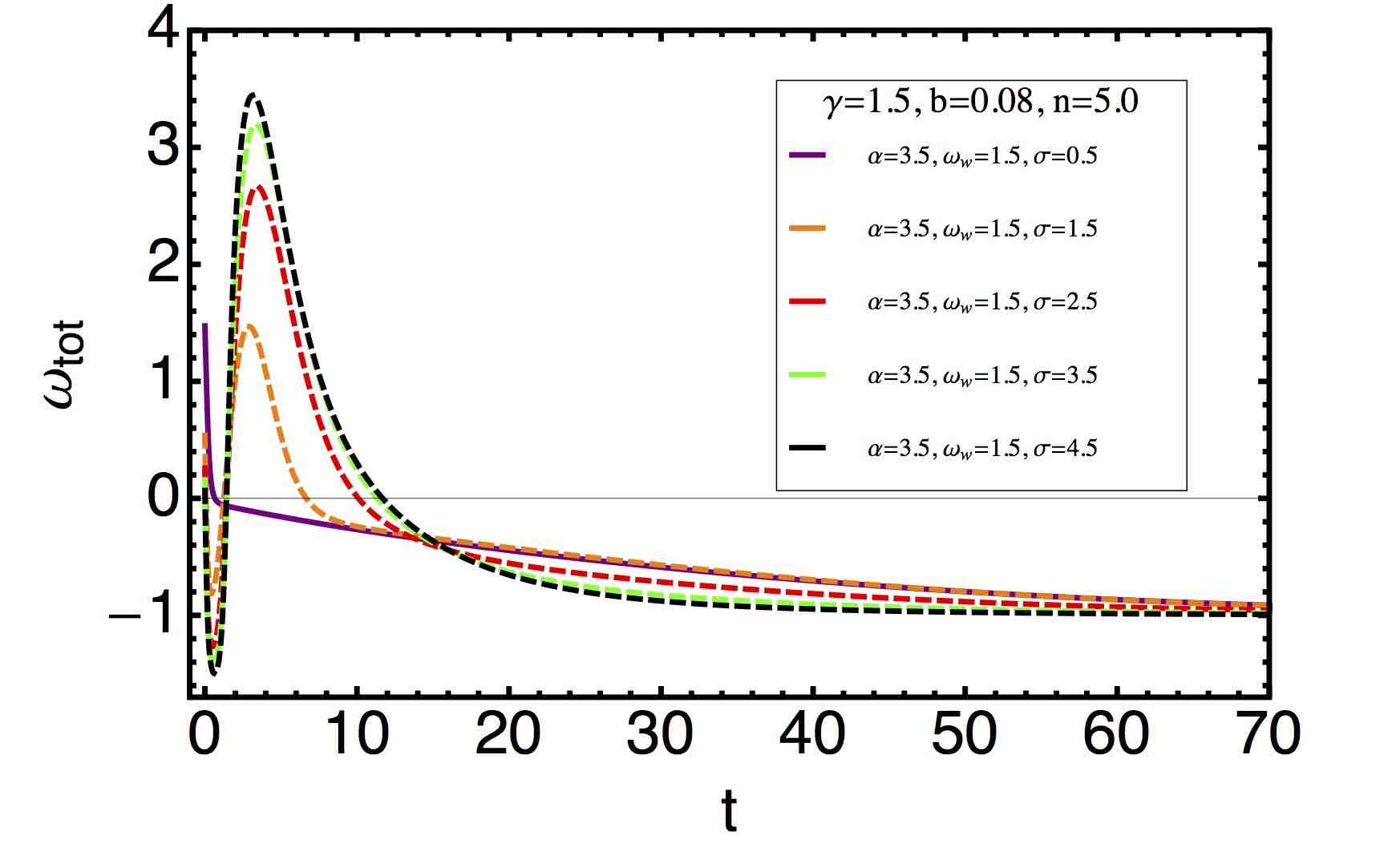} &
\includegraphics[width=50 mm]{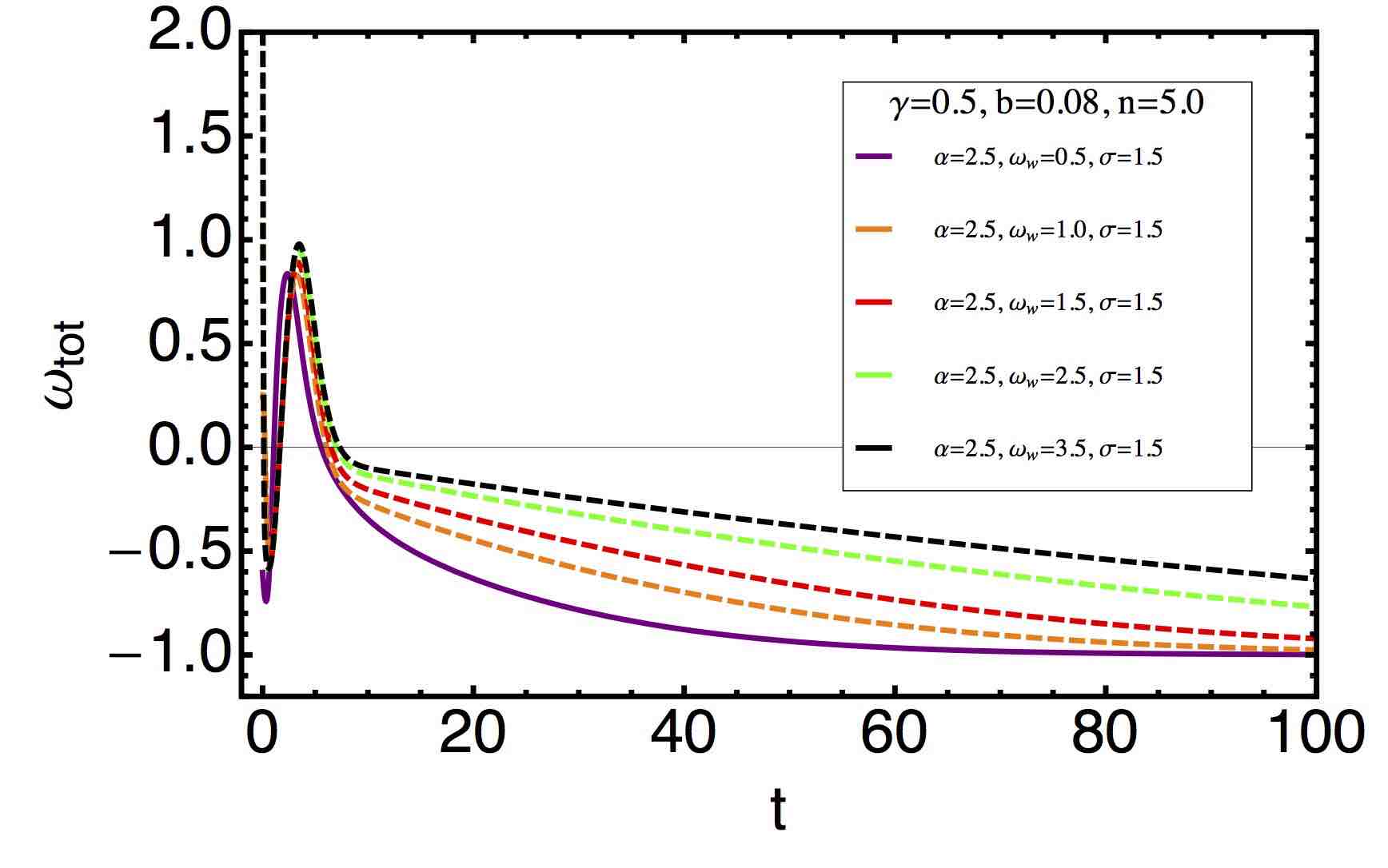}
 \end{array}$
 \end{center}
\caption{Behavior of scale factor $\omega_{tot}$ against $t$.}
 \label{fig:20}
\end{figure}

\begin{figure}[h!]
 \begin{center}$
 \begin{array}{cccc}
\includegraphics[width=50 mm]{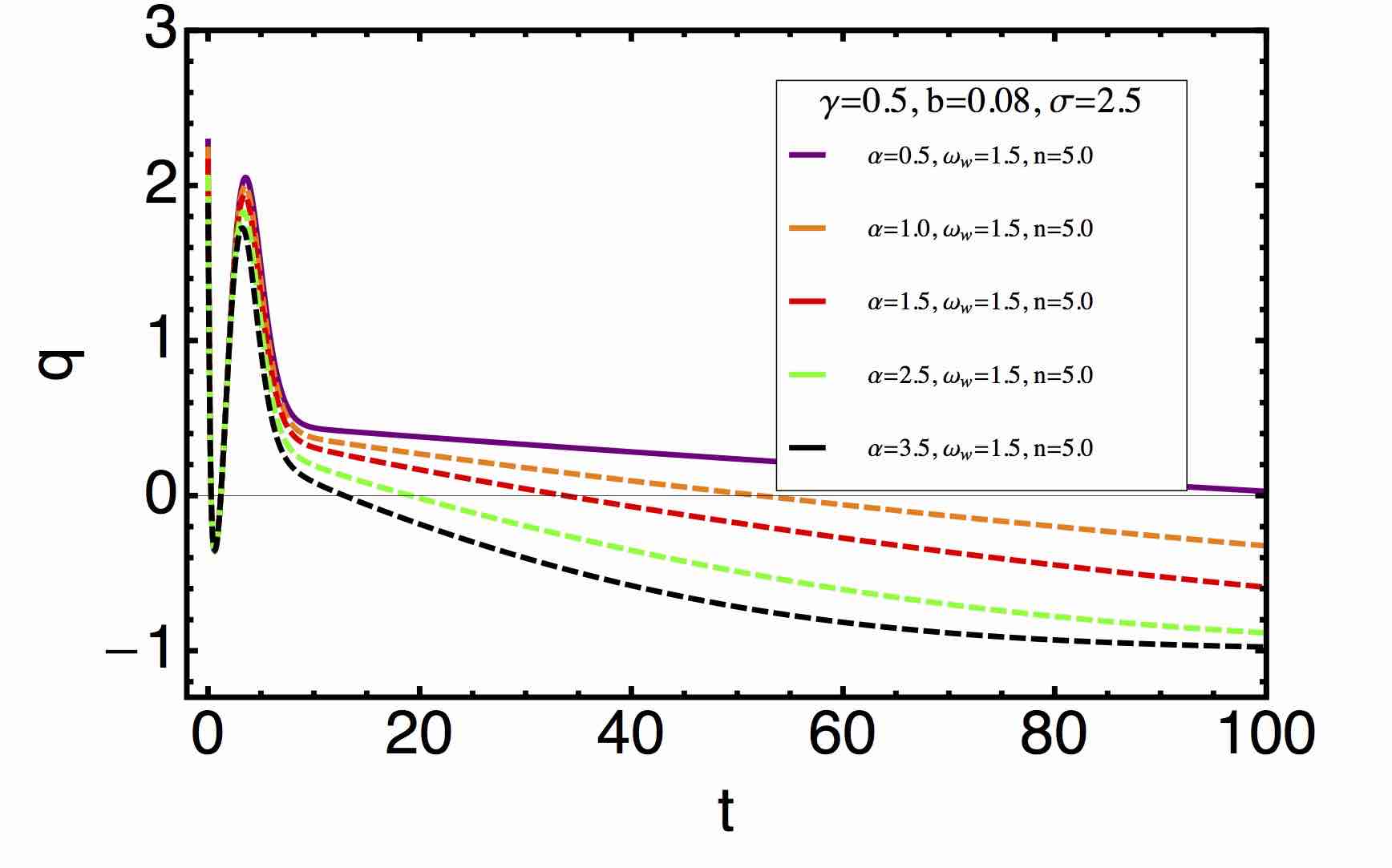} &
\includegraphics[width=50 mm]{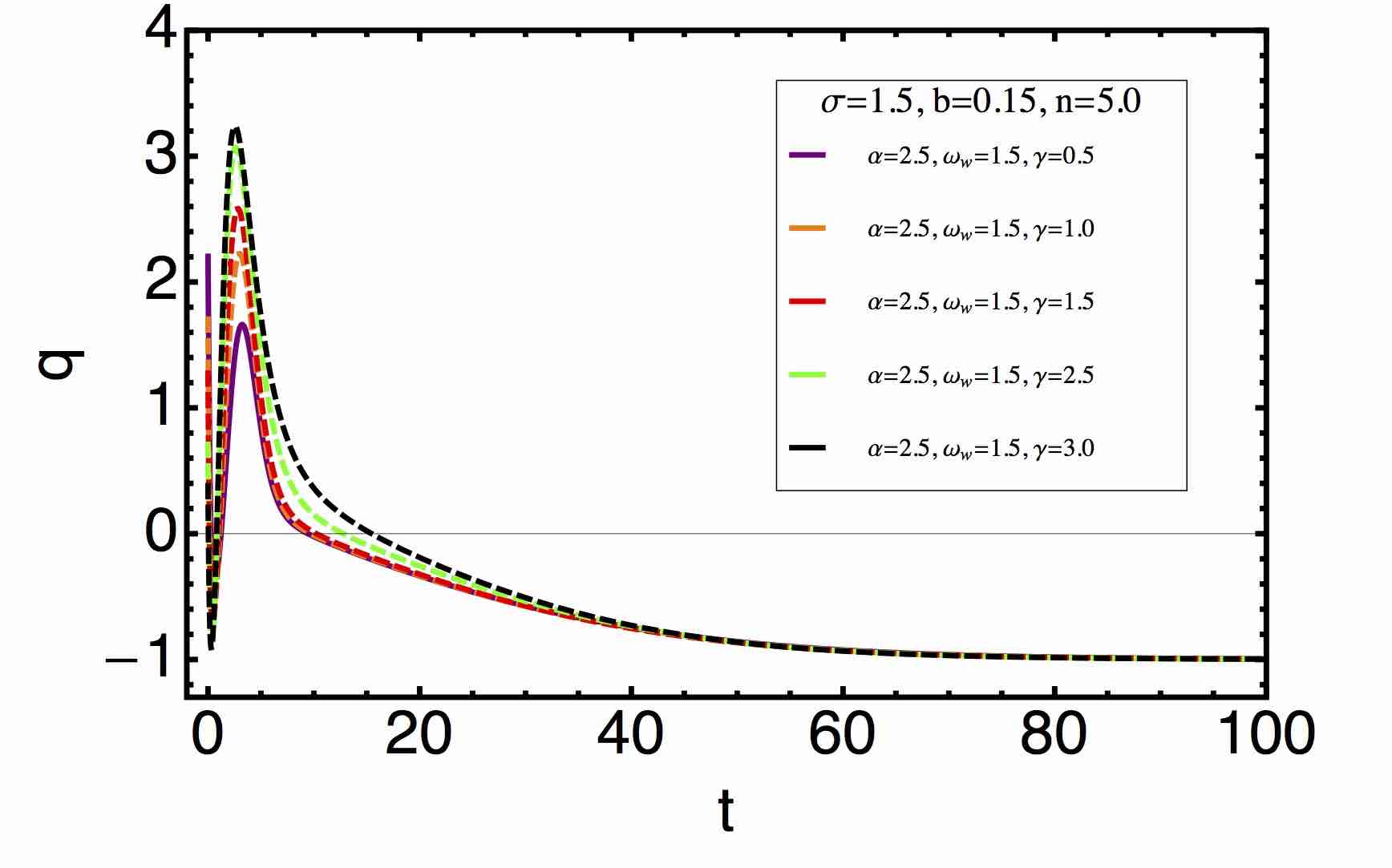}\\
\includegraphics[width=50 mm]{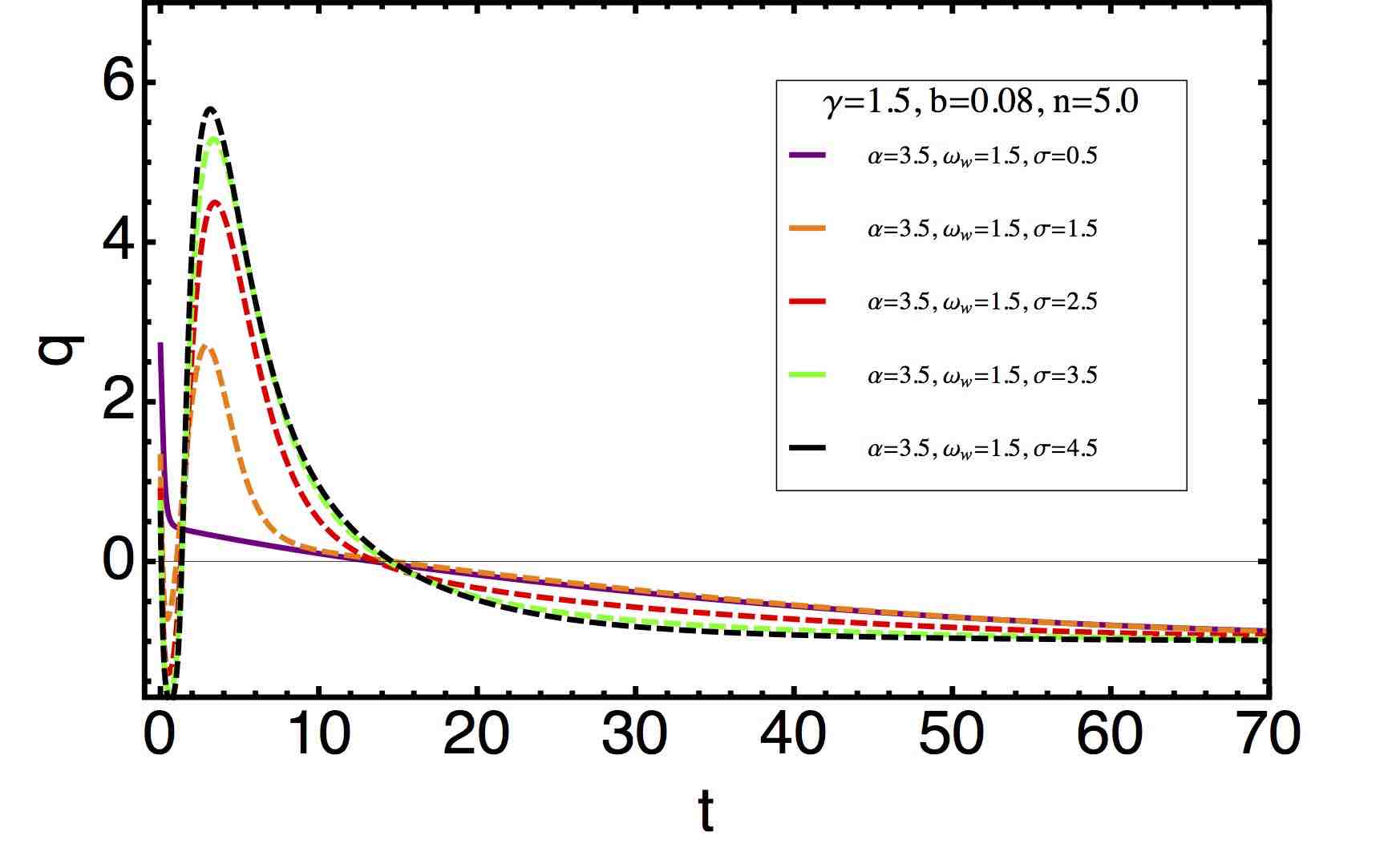} &
\includegraphics[width=50 mm]{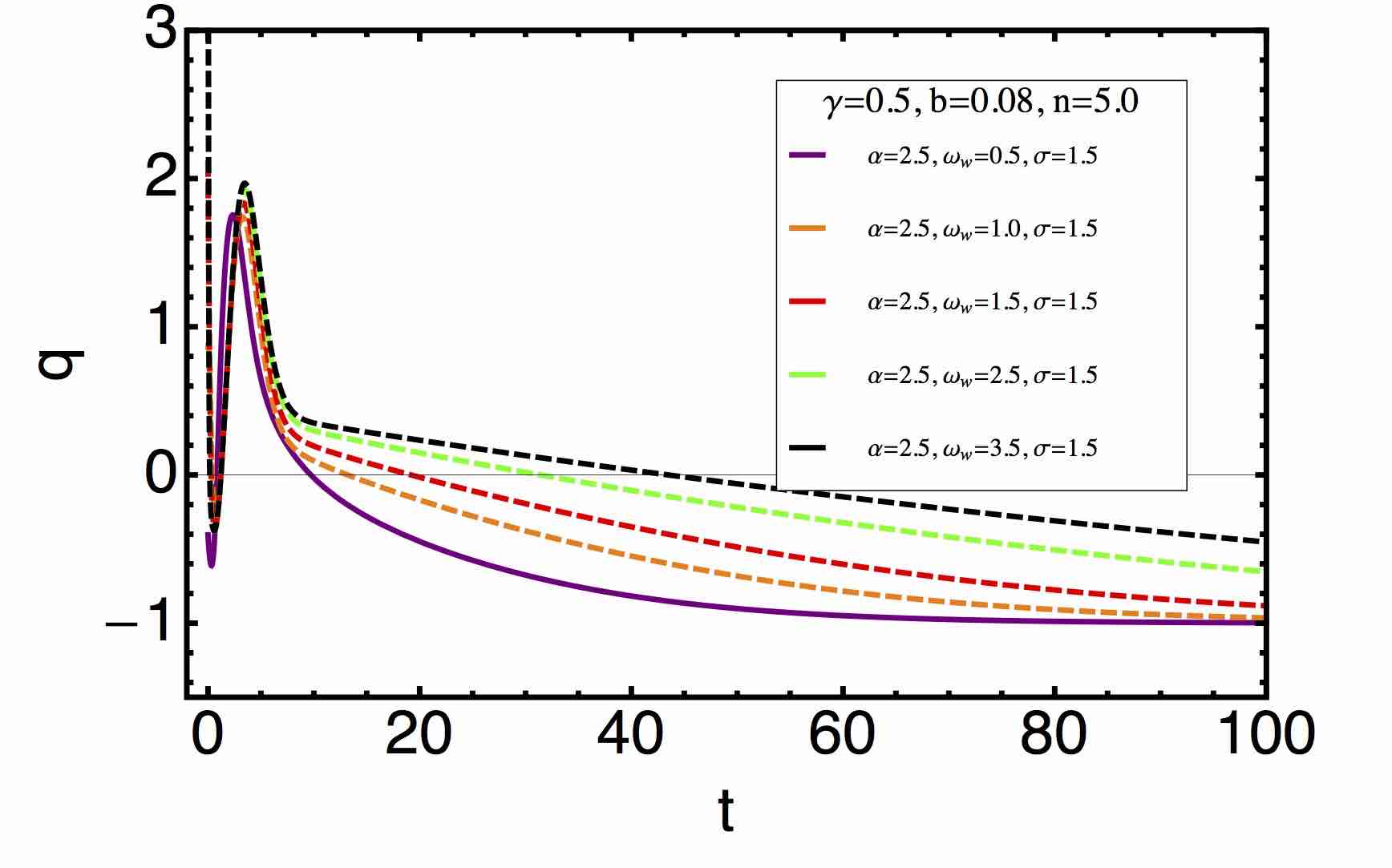}
 \end{array}$
 \end{center}
\caption{Behavior of deceleration parameter $q$ against $t$.}
 \label{fig:30}
\end{figure}

\section{Discussion}
We suppose a model of a Universe consists of generalized ghost dark energy, Van der Waals fluid interacting with a modified fluid, which could be result of some interactions (we could only one of the possible forms). Interaction between components were assumed to be $Q=3Hb(\rho_{\small{tot}}-\rho_{GDe})$. By this way the role of the generalized ghost dark energy were seen by interaction term only. During the investigation we recover scale factor $a(t)$ and present graphical analysis of $\ddot{a}$  and EoS parameter $\omega_{\small{tot}}$. Analysis shows that the effects concerning to the presence of a generalized ghost dark energy expressed in the interaction term was observed in later stages of the evolution comparing with cases corresponding to absence and presence of an interaction of the form $Q=3bH\rho_{\small{tot}}$ between components. From the graphical analysis we see that having a situation proposed in this article is not always true the assumption that dark energy could be the source of the accelerated expansion of the Universe. At least, for some values of the parameters of the model we observe that the mixture behaves as a dark energy with negative EoS parameter, while $\ddot{a}<0$. For some cases we saw that dark energy really is responsible for accelerated expansion with $\ddot{a}>0$. Almost the same result we obtain considering a modified fluid not described by the energy density of our consideration, but as a fluid described as $P=-\rho$. Which corresponds to a cosmological constant model with $\omega=-1$. Here, we have following possibilities. First, the modification of a fluid of our consideration could not work with Van der Waals gas, or probably an interaction of the consideration between components is not realistic and other modification should be done with interaction term. There is other possibility, which from our opinion is also has right to be, that dark energy is not always was and will be responsible for the accelerated expansion of the Universe.\\\\\\

\section*{\large{Appendix A: The Generalized Second Law of Thermodynamics}}
In this section we are going to deal with the question of the validity of the Generalized Second Law (GSL) of thermodynamics. For GSL of thermodynamics we will follow \cite{Ujjal} for the writings of this section (references therein are useful for complete introduction), where was considered validity of the GSL of thermodynamics for the Universe bounded by the Hubble horizon\footnote{
Recall that in case when $k=0$ as in our case apparent horizon
$R_{A}=\frac{1}{\sqrt{H^{2}+\frac{k}{a^{2}}}}$
we get the radius of the Hubble horizon (\ref{eq:Habblehor}).}

\begin{equation}\label{eq:Habblehor}
R_{H}=\frac{1}{H},
\end{equation}
cosmological event horizon,
\begin{equation}\label{eq:cosevhor}
R_{E}= a\int_{t}^{\infty}\frac{dt}{a},
\end{equation}
and the particle horizon,
\begin{equation}\label{eq:particlehor}
R_{P}=a\int_{0}^{t}\frac{dt}{a}.
\end{equation}
The contents in the Universe bounded by the event horizons taken as interacting two components of a single scalar field. The foundation of GSL required the Gibb’s equation of thermodynamics is,
\begin{equation}\label{Gibs}
  T_{X}dS_{IX}= PdV_{X} + dE_{IX}
\end{equation}
where $S_{IX}$ and $E_{IX}=\rho V_{X}$, are  internal entropy and energy within the horizon, while $V_{X}=\frac{4}{3}\pi R^{3}_{X}$ be the volume of sphere with horizon radius \[R_{X}=\left(\sqrt{H^{2}+\frac{k}{a^{2}}}\right)^{-1}. \].\\
Recall that GSL with first law for the time derivative of total entropy gives,
\begin{equation}\label{eq:UF}
\dot{S}_{X}+\dot{S}_{IX}=\frac{R^{2}_{X}}{GT_{X}}\left(\frac{k}{a^{2}}-\dot{H}\right)\dot{R}_{X},
\end{equation}
while in case without the first law used we get,
\begin{equation}\label{eq:UNUF}
\dot{S}_{X}+\dot{S}_{IX}= \frac{2\pi R_{X}}{G}\left[ R^{2}_{X}\left(\frac{k}{a^{2}}-\dot{H}\right)(\dot{R}_{X}-HR_{X})+\dot{R}_{X} \right].
\end{equation}
Under the notations used above we understood that $T_{X}=\frac{1}{2\pi R_{X} }$ and $R_{X}$ is temperature and Radius for a given horizon under equilibrium thermodynamics respectively, $S_{X}$ is the horizon entropy and $\dot{S}_{IX}$ as the rate of change of internal entropy. It was found that the first and second laws of thermodynamics hold on the apparent horizon when  the apparent horizon and the event horizon of the Universe are different, while for consideration of only event horizon these laws breakdown \cite{Wang}. The Friedmann equations and the first law of thermodynamics ( on the apparent horizon ) are equivalent if the Universe is bounded by the apparent horizon $R_{A}$ with temperature $T_{A}=\frac{1}{2\pi R_{A}}$ and entropy $S_{A}=\frac{\pi R_{A}}{G}$\cite{Cai3}.Usually, the Universe bounded by apparent horizon and in this region the Bekenstein's entropy - mass bound ($S \leq 2\pi E R_{A}$) and entropy - area bound ($S \leq\frac{A}{4}$) are hold.\\
In order the GSL to be hold it is required that $\dot{S}_{X}+\dot{S}_{IX}\geq0$ i.e. the sum of entropy of matter enclosed by horizon must be not be a decreasing function of time.

\section*{\large{Appendix B: Fluid with general $f(a,n)$}}
Our interest was a model where following form for $f(a)$ could be considered
\begin{equation}\label{eq:fmod000}
f(a)=1+\gamma a^{n}\exp[-a^{2}/\sigma^{2}].
\end{equation}
Numerical analysis for the model were performed to illustrate behavior of the Universe with new factor $f(a,n)$. We follow satisfy GSL and obtained following behavior for cosmological parameters. We can see that presence of interaction is necessary to obtain negative pressure and positive density at the late time (see Fig. 12). Also we can see from Fig. 13 expected behavior of the deceleration and total EoS parameters.
\begin{figure}[h!]
 \begin{center}$
 \begin{array}{cccc}
\includegraphics[width=50 mm]{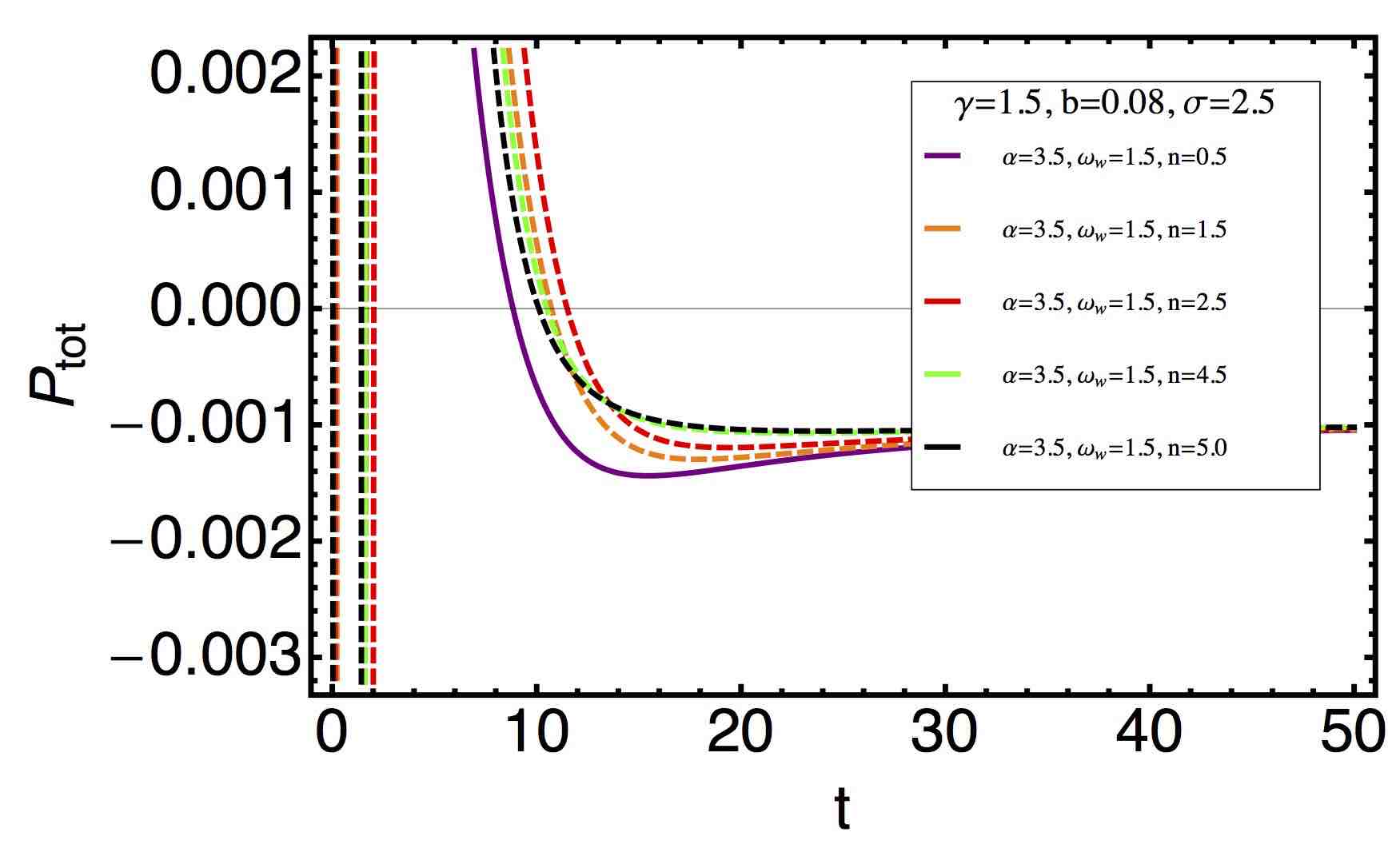} &
\includegraphics[width=50 mm]{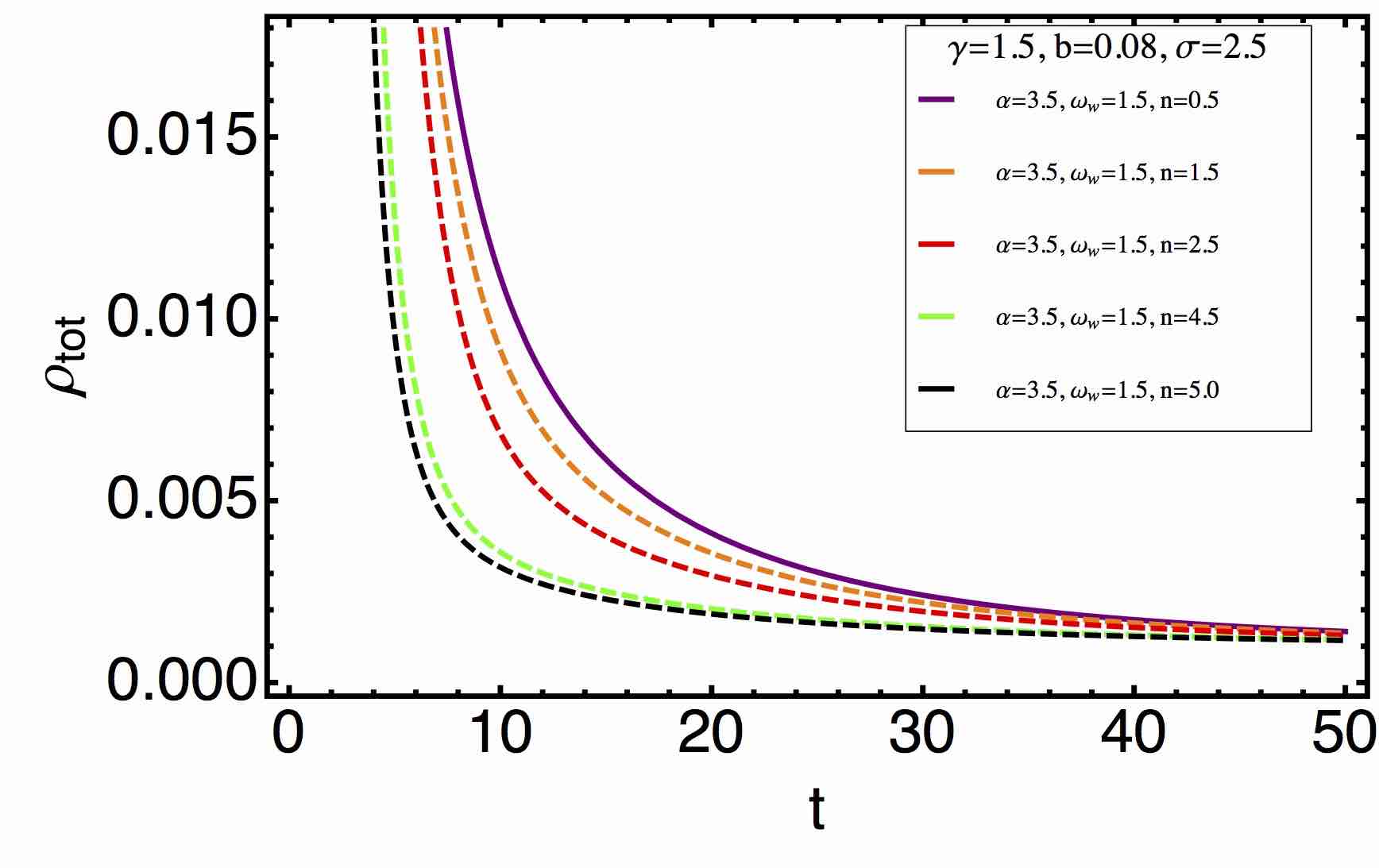}\\
\includegraphics[width=50 mm]{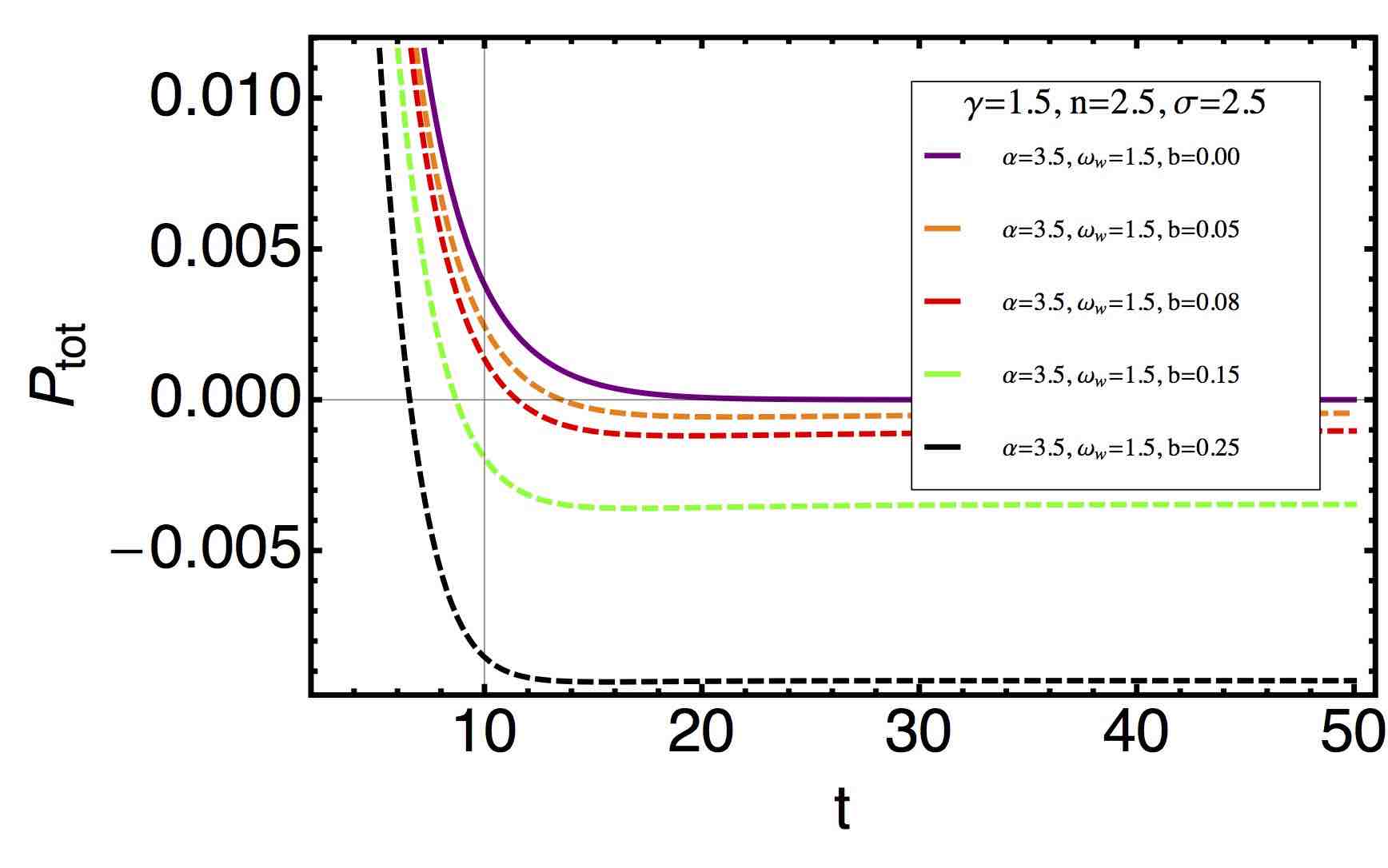} &
\includegraphics[width=50 mm]{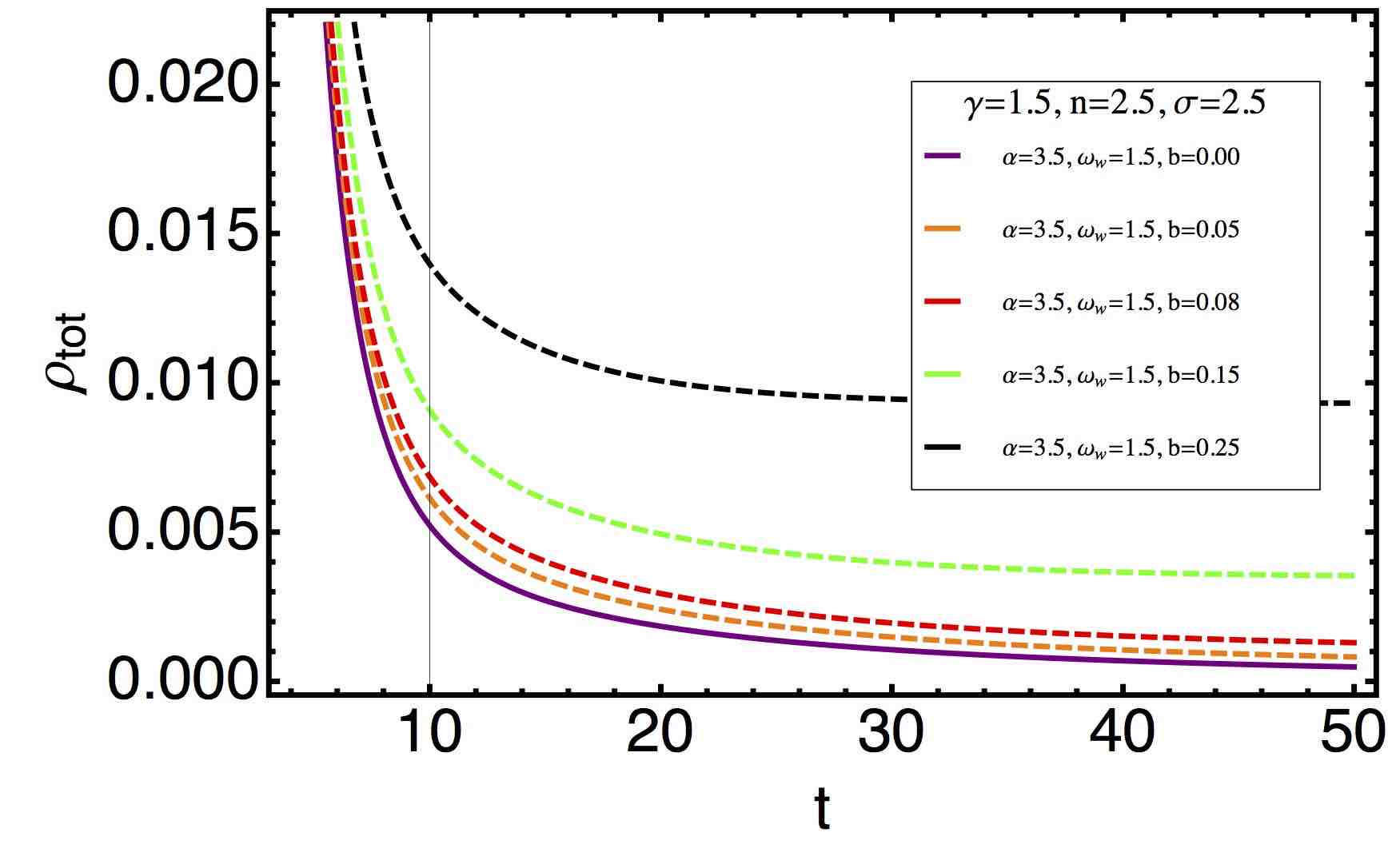}
 \end{array}$
 \end{center}
\caption{Behavior of deceleration parameter $q$ against $t$ for different values of interaction parameter $b$ for $f(a,n)$ form.}
 \label{fig:5}
\end{figure}
\vspace{100 mm}
\begin{figure}[h!]
 \begin{center}$
 \begin{array}{cccc}
\includegraphics[width=50 mm]{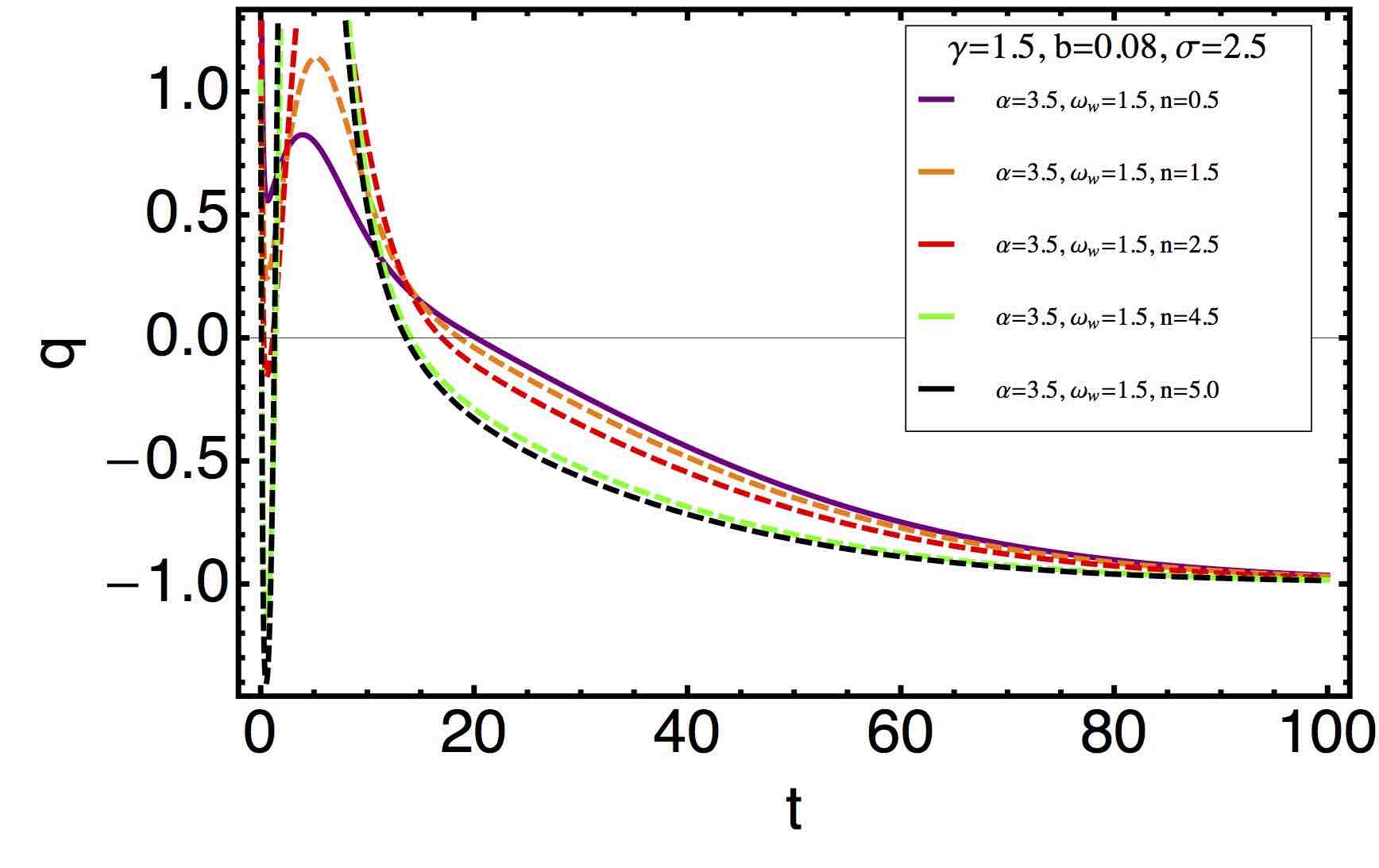} &
\includegraphics[width=50 mm]{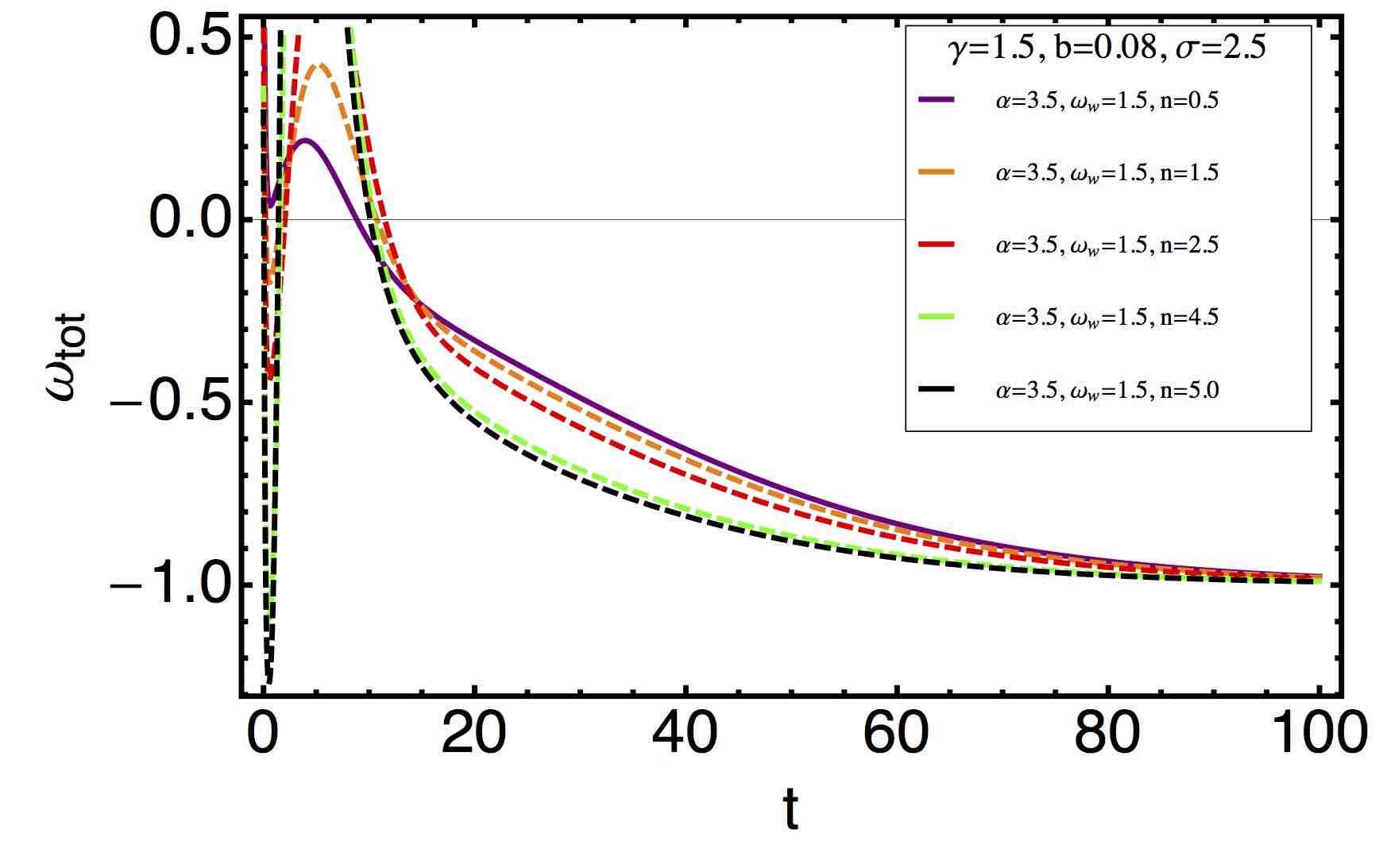}\\
\includegraphics[width=50 mm]{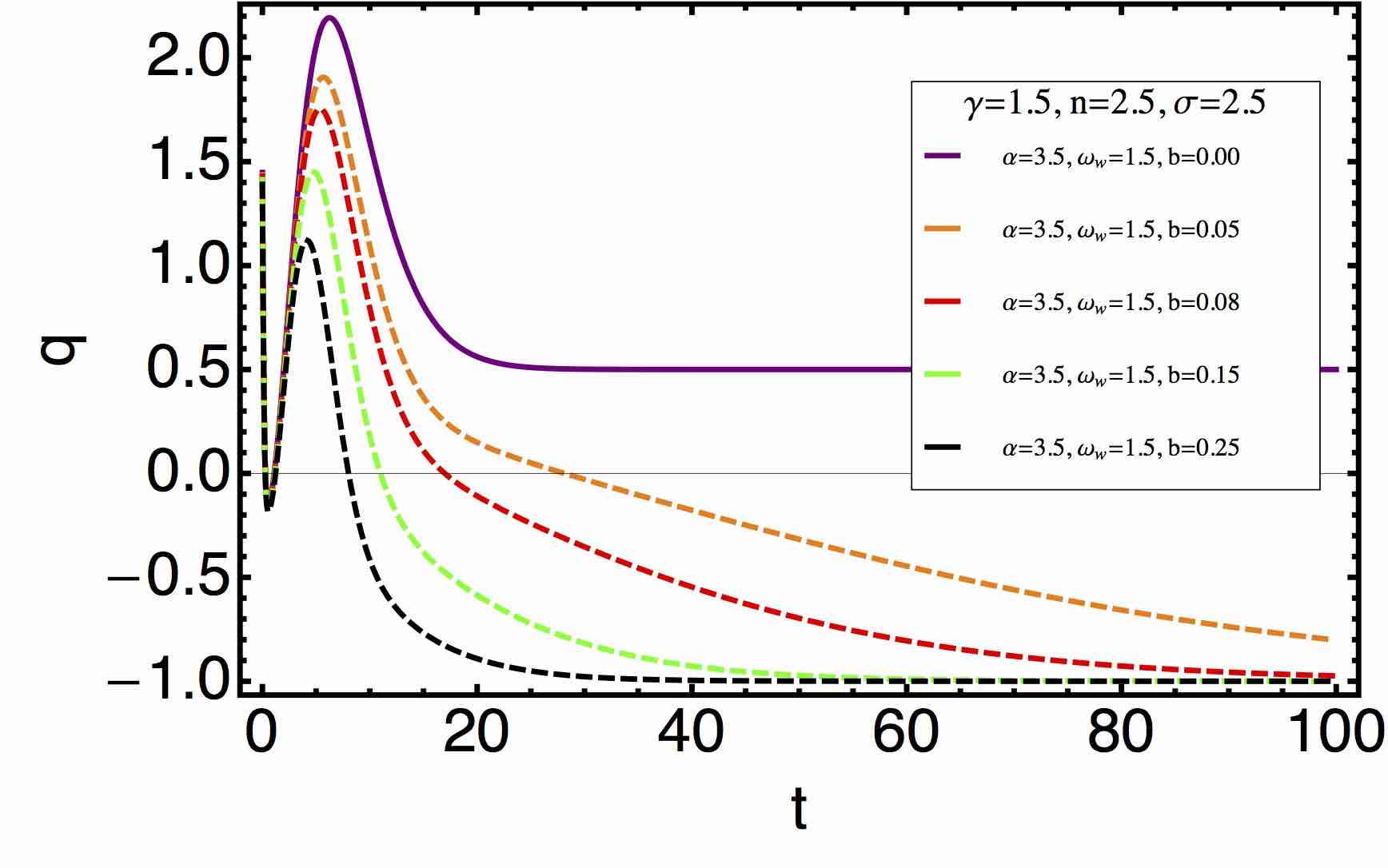} &
\includegraphics[width=50 mm]{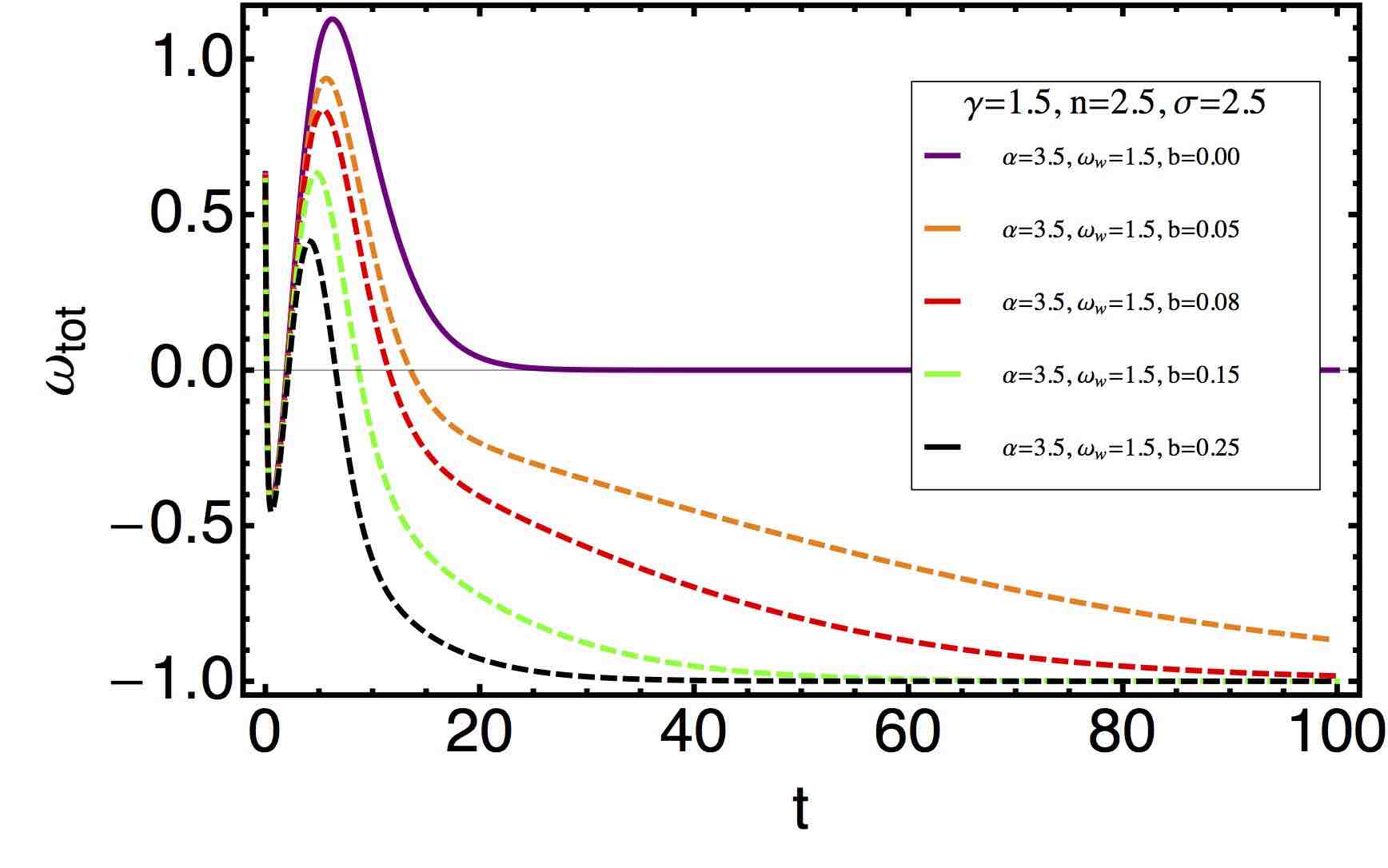}
 \end{array}$
 \end{center}
\caption{Behavior of deceleration parameter $q$ against $t$ for different values of interaction parameter $b$ for $f(a,n)$ form.}
 \label{fig:6}
\end{figure}

\section*{\large{Appendix C: Interaction term}}
\begin{figure}[h!]
 \begin{center}$
 \begin{array}{cccc}
\includegraphics[width=50 mm]{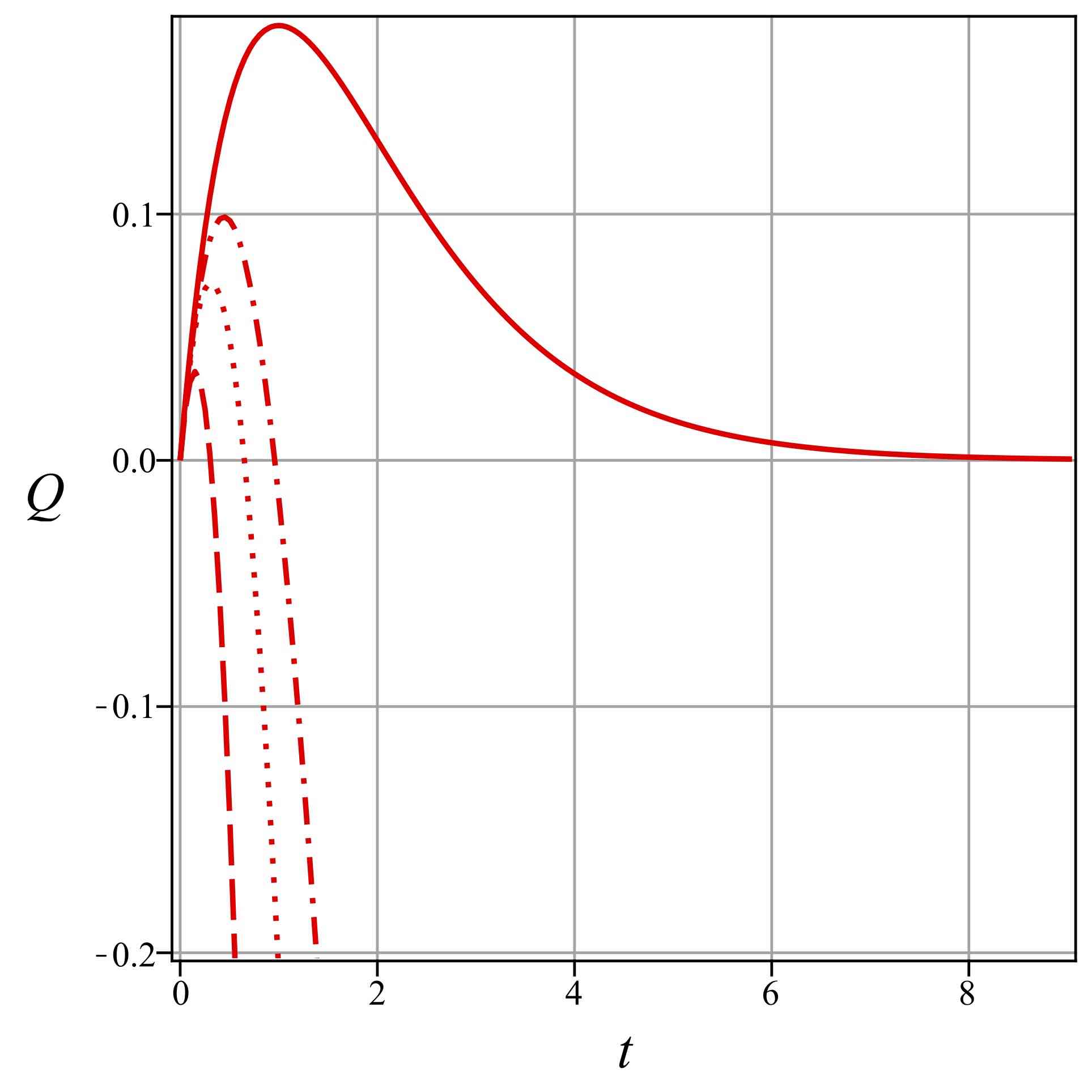} &
\includegraphics[width=50 mm]{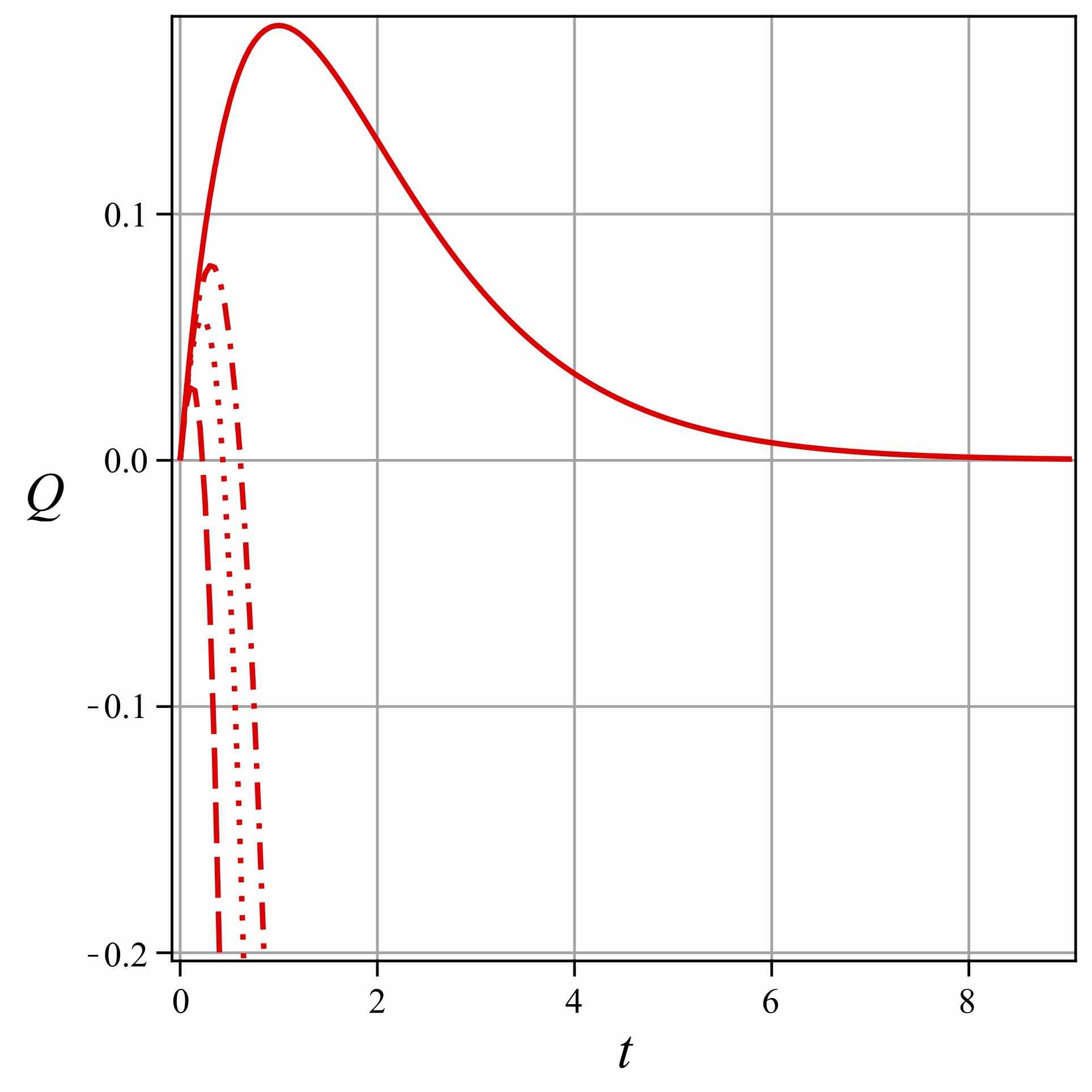}
 \end{array}$
 \end{center}
\caption{Behavior of interaction term $Q$ against $t$ for different values of $\alpha$ for $b=0.08$. Left: Model of the ghost dark energy with $\alpha=0$ (solid), $\alpha=0.2$ (dash dot), $\alpha=0.4$ (dot), $\alpha=1.2$ (dash). Right: Model of the generalized ghost dark energy with $\alpha=0$ (solid), $\alpha=0.2$ (dash dot), $\alpha=0.4$ (dot), $\alpha=1.2$ (dash).}
 \label{fig:Q1}
\end{figure}

Having Hubble parameter and energy densities let us to discuss about interaction term,
\begin{equation}\label{end}
Q=3Hb(\rho_{\small{tot}}-\rho_{GDe})
\end{equation}
and possible the value of the coupling constant $b$. Our stability analysis suggested $b<0.3$ and we found the best fitted value as $b\approx0.08$. Now we can give plot of interaction term in terms of cosmic time. From the Fig. \ref{fig:Q1} we can see that interaction term yields to negative value at the late time for non-zero $\alpha$ and $\beta$. However at the early universe we have positive interaction term. In the case of zero $\alpha$ and $\beta$ interaction term vanish at the late time. On the other hand for the case of non-zero $\alpha$ and $\beta$ at the late time the interaction term has large value which means strong coupling.

\end{document}